\documentclass[lettersize,journal]{IEEEtran}
\pdfoutput=1
\usepackage{amsmath,amsfonts}
\usepackage{algorithmic}
\usepackage{algorithm}
\usepackage{array}
\usepackage{hyperref}
\usepackage{cleveref}
\usepackage{subfig}
\usepackage{textcomp}
\usepackage{stfloats}
\usepackage{url}
\usepackage{color}
\usepackage{verbatim}
\usepackage{graphicx}
\usepackage{cite}
\usepackage{bm}
\usepackage{amssymb}

\begin{document}

\title{Multi-Person Passive WiFi Indoor Localization with Intelligent Reflecting Surface}

\author{Ganlin Zhang, Dongheng Zhang, Ying He, Jinbo Chen, Fang Zhou 
        \\ and Yan Chen
\IEEEcompsocitemizethanks{\IEEEcompsocthanksitem Ganlin Zhang, Dongheng Zhang, Jinbo Chen, Fang Zhou, Yan Chen are with the school of Cyber Science and Technology, University of Science and Technology of China, Hefei 230026, China (E-mail: zgl1362@mail.ustc.edu.cn, dongheng@ustc.edu.cn, jinbochen@mail.ustc.edu.cn, fangzhou1958@mail.ustc.edu.cn, eecyan@ustc.edu.cn).

Ying He is with the School of Information and Communication, University of Electronic Science and Technology of China, Chengdu 611731, China (email: heying@std.uestc.edu.cn).\protect\\
}}



\maketitle

\begin{abstract}
The past years have witnessed increasing research interest in achieving passive human localization with commodity WiFi devices. However, due to the fundamental limited spatial resolution of WiFi signals, it is still very difficult to achieve accurate localization with existing commodity WiFi devices. To tackle this problem, in this paper, we propose to exploit the degree of freedom provided by the Intelligent Reflecting Surface (IRS), which is composed of a large number of controllable reflective elements, to modulate the spatial distribution of WiFi signals and thus break down the spatial resolution limitation of WiFi signals to achieve accurate localization. Specifically, in the single-person scenario, we derive the closed-form solution to optimally control the phase shift of the IRS elements. In the multi-person scenario, we propose a Side-lobe Cancellation Algorithm to eliminate the near-far effect to achieve accurate localization of multiple persons in an iterative manner. Extensive simulation results demonstrate that without any change to the existing WiFi infrastructure, the proposed framework can locate multiple moving persons passively with sub-centimeter accuracy under multipath interference and random noise.  

\end{abstract}

\begin{IEEEkeywords}
Intelligent Reflecting Surface; Passive Indoor Localization; Device-free; WiFi Localization; Near-far Effect. 
\end{IEEEkeywords}

\section{Introduction}
\IEEEPARstart{P}assive indoor localization without any device attached to the human has received great attention recently, which supports a growing list of location-based services and applications, including intrusion detection \cite{lv2019robust, wang2019wi}, elderly monitoring \cite{hong2016socialprobe, husen2014indoor}, and real-time positioning of criminals \cite{tundis2020detecting}. Vision-based technologies \cite{werner2011indoor, wolf2005robust} are limited by lighting conditions and privacy concerns for real-life deployment, while radar equipment \cite{chen2020speednet, zhang2020mtrack, zhang2018multitarget, chen2019residual} is too expensive for large-scale deployment. Due to the ubiquitous availability of WiFi devices, a WiFi-based system~\cite{zhang2019calibrating, kumar2014accurate, han2016enabling} is a cost-effective solution. 

Existing WiFi-based methods mainly rely on the estimation of angle-of-arrival (AoA) and/or time-of-flight (ToF). However, due to the limited bandwidth and the number of antennas of the commodity WiFi devices, the resolution of the estimated AoA and ToF is generally poor. In such a case, the passive localization accuracy is poor, and thus the applications are limited. To resolve the resolution limitation of the commodity WiFi devices, one possible solution is to utilize the recently emerging Intelligent Reflecting Surface (IRS)~\cite{tang2020wireless, wu2021intelligent, li2017electromagnetic}.


IRS is a digitally controlled metasurface consisting of a large number of small passive elements, each of which can independently tune the phase control of the incident signals by changing the applied control voltage. 
Different from conventional phased arrays which require high power consumption and implementation cost, IRS has much lower power consumption and implementation cost by passively reflecting the incoming signals. 
Furthermore, IRS possesses other advantages such as full-duplex mode and flexible deployment characteristics. 
By directly reshaping the signal reflected by all the IRS elements, e.g., constructively superimposing with signals at a specific location or destructively interfering with signals at other locations, IRS can reconfigure the wireless signals to illuminate the scene. 
With a large number of low-cost controllable elements, IRS provides a new degree of freedom that makes it possible to break through the resolution limitation of commodity WiFi devices without making any changes to existing infrastructures.  

Nevertheless, passive human localization with the aid of IRS is not trivial since the reflected signals from the targeted person are very likely to be submerged in the interference signals reflected from other reflectors. Thus, to achieve satisfying localization performance, several challenges are needed to be resolved.


\textbf{How to deal with the multipath effect caused by static reflectors?} 
Multipath interference is a fundamental and challenging problem in wireless systems, which becomes more troublesome for passive human sensing where human reflections are much weaker than multipath. To resolve this problem, we have noted that multipath keeps invariant in the time domain while human reflections vary over time, which makes it feasible to handle multipath leveraging such discrepancy in the time domain.  

\textbf{How to extract the signals from a specific location?} 
The limited spatial resolution of commodity WiFi devices makes it difficult to extract the signals from a specific location, which further limits the localization accuracy.  
To resolve this challenge, we have noted that IRS could be composed of a large number of elements, which could provide sufficient spatial resolution for accurate localization.
By searching the optimal phase shift of IRS elements, we could enhance the signals from the specific locations while suppressing other signals.

\textbf{How to handle the near-far effect in the multi-person scenario?} 
In practical indoor environments, the signal reflected by the nearest person can be much stronger than that of distant persons, making it hard to localize the distant persons with weaker reflection signals. To resolve this challenge, we propose a Side-lobe Cancellation algorithm by optimally controlling the IRS elements to strengthen the signals reflected from the distant person while weakening that from the detected persons.  

Overall, the contributions of this article can be summarized as follows: 
\begin{itemize}
\item To the best of our knowledge, this paper is the first attempt towards utilizing IRS for passive indoor localization. By creating fine-grained reflected beams through optimally controlling the IRS, we break through the spatial resolution limitation of commodity WiFi devices.
\item In the single-person scenario, we formulate the IRS optimization problem by maximizing the signal amplitude difference between the moving person and the multipath interference and derive the closed-form solution to control the phase of the IRS elements. In the multi-person scenario, we propose a Side-lobe Cancellation algorithm to resolve the near-far effect to achieve accurate localization. 
\item Extensive simulations have been conducted to demonstrate that the proposed method can accurately localize the moving persons even in the presence of multi-person, multipath interference, and severe random noise.  
\end{itemize}

The rest of this paper is organized as follows. Section~\ref{Related_Work} describes the related work. In Section~\ref{Model}, we introduce the model of our system. In Section~\ref{section_Single_Person} and Section~\ref{section_Multi_Person}, we present the optimization problems and the solutions for single-person and multi-person scenarios, respectively. The simulation results are shown in Section~\ref{Simulation}. Finally, the conclusion is drawn in Section~\ref{Conclusion}.

The notations adopted in this paper are listed in Table~\ref{table_notion}.

\section{Related Work}
\label{Related_Work}
In this section, we discuss the related work on passive indoor localization, IRS, and wireless sensing, respectively. 
\subsection{Passive Indoor Localization}
Passive indoor localization with WiFi can be generally categorized into two types: RSSI (Received Signal Strength Indicator) based methods and CSI (Channel State Information) based ones.
The RSSI-based technology realizes localization by characterizing the energy attenuation of wireless signals during propagation, which is readily available from commodity WiFi.
Several methods combine the RSSI readings from multiple access points (APs) with propagation models and then locate the persons through triangulation~\cite{schmidt1986multiple} or fingerprint collection~\cite{sun2016indoor, duan2019data}. 
However, since RSSI is a coarse characterization of the channel, reliable RSSI-based positioning systems tend to achieve only room-level accuracy. 
Compared with the RSSI, CSI~\cite{zhang2018breath, vasisht2016decimeter, li2016dynamic, wang2018low} is a fine-grained characterization of the channel that can provide not only the amplitude but also the phase information.  
\cite{zhang2019breathtrack} exploits the phase variation of the CSI and proposes a joint AoA-ToF sparse recovery method to extract the information of the dominant path for localization. \cite{qian2018widar2} jointly considers the AoA, ToF, and Doppler shift and designs an efficient estimation algorithm. However, due to the limitation of the number of antennas and the bandwidth of the signal, the accuracy and robustness of these systems are still limited. For instance, the median localization accuracy of \cite{qian2018widar2} and \cite{li2016dynamic} is 75cm and 60cm respectively, which might place the person in a different room within a building. To achieve a higher location accuracy, \cite{xie2019md} requires eight antennas for indoor localization whereas a typical WiFi device only has three, and thus hardware modifications are needed which limits its applications. 

\begin{table}
\begin{center}
\caption{Notations adopted in this paper.}
\label{table_notion}
\begin{tabular}{ l  l }
\hline
\textbf{Notation} & \textbf{Description}\\
\hline
$f$ & the carrier frequency in Hertz (Hz)\\
$\lambda$ & the wavelength of the transmitted signal\\
$s_{tx}$ & the equivalent complex-valued transmit signal \\
$N_r$ & the number of receiving antenna\\
$M$ & the number of the IRS element\\
$P$ & the transmit power \\
$\bm{h_{TO}}$ & the channel vector from the transmitter to the reflector\\
$\bm{h_{OR}}$ & the channel vector from the reflector to the receiver\\
$\bm{h_{TI}}$ & the channel vector between the IRS and the reflector \\
$\bm{h_{IO}}$ & the channel vector between the transmitter and the IRS \\
$\bm{q}$ & the weighting vector of the IRS\\
$\bm{w}$ & the beamforming vector of the receiver\\
$\bm{\epsilon}$ & the circularly symmetric complex Gaussian noise vector\\
$\rho_0$ & the path loss at the reference distance $d_0 = 1 \rm{m}$\\
$\alpha$ & the path loss exponent\\
$a$ & a scalar\\ 
$\bm{a}$ & a vector \\
${a}^{m}$ & $m$-th element of vector $\bm{a}$ \\
$\bm{A}$ & a matrix \\
${A}^{m,n}$ & the element in $m$-th row and $n$-th column  \\
& of matrix $\bm{A}$\\
diag\{$\bm{a}$\} & the diagonal matrix with the elements of vector $\bm{a}$ \\
&on its main diagonal\\
$\cdot$ & the Hadamard product \\
$\times$ & the matrix product \\
$|a|$ & $l_{1}$ norm of $a$\\
$\bm{A}^{T}$ & the transpose matrix of $\bm{A}$\\
$\bm{A}^{H}$ & the conjugate matrix of $\bm{A}$\\
$\text{tr}(\bm{A})$ & the trace of matrix $\bm{A}$\\
$\text{rank}(\bm{A})$ & the rank of matrix $\bm{A}$\\
\hline 
\end{tabular}
\end{center}
\end{table}

\subsection{Intelligent Reflecting Surface}
IRS is recently developed as a promising new paradigm to customize the radio environment and enhance the wireless link quality for 5G/6G wireless communication. It has shown potential advantages in many applications, including enhanced wireless communication capacity~\cite{wu2020joint, tao2020performance, wang2020energy}, secure transmission~\cite{cui2019secure, chu2019intelligent}, and coverage extension~\cite{wu2019towards, zhang2020beyond}. More recently, several works~\cite{zhang2021metalocalization,fascista2021ris, he2020large, hu2018beyond} have been discussed by using IRS for localization. \cite{fascista2021ris, he2020large, hu2018beyond} consider outdoor scenarios and utilize a large-scale base station to transmit signals for localization, which is not suitable for indoor scenarios. \cite{zhang2021metalocalization} utilizes the IRS for indoor localization by comparing the actual RSSI measured by the mobile device with the theoretical value of each location in space. However, this RSSI-based method does not take the multipath problem into account, which may lead to unstable performance when applied in real multipath environments. Furthermore, these existing works focus on active positioning which means that the person needs to carry a mobile device to actively search and collect signals. For passive human localization systems where persons do not carry any active devices for signal transmission, the reflection from the person can be much weaker than that from the other reflectors. Therefore, achieving accurate passive indoor localization is still an unresolved problem. 

\subsection{Wireless Sensing}
Benefitting from the non-contact nature of wireless signals, human sensing based on wireless signals has attracted great interest in the past decades. 
Many applications including gesture recognition~\cite{li2021towards, zhang2021unsupervised, regani2021wifi}, vital signs monitoring~\cite{adib2015smart, katabi2014tracking, chen2021contactless}, human activity imaging~\cite{he2020non, karanam20173d} have been achieved.  
The foundation of these applications lies in the fact that the movement of the human in the environment would modulate the propagation of wireless signal, which makes it feasible to extract human information from the signal variation~\cite{stove1992linear}. 

Different from existing works, in this paper, we investigate high-accurate passive indoor localization with commodity WiFi devices by exploiting the degree of freedom provided by the IRS. 


\begin{figure*}[tb]
\centering
\includegraphics[width=0.8\textwidth]{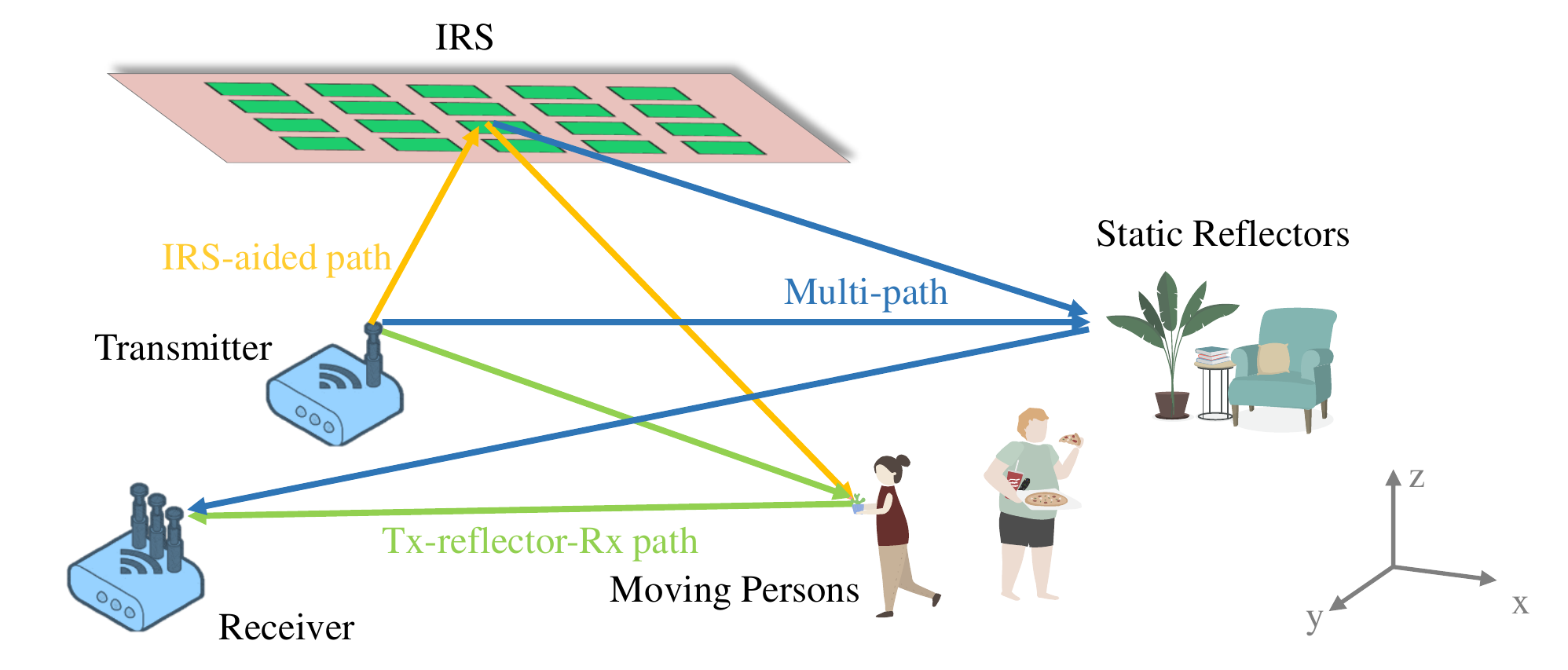}
\caption{An illustration of the multi-person passive localization system with IRS.}
\label{IRS_picture}
\end{figure*}

\section{System Model}
\label{Model}
As shown in Fig.~\ref{IRS_picture}, our system is composed of a single-antenna transmitter (Tx), a receiver (Rx) with $N_{r}$ antennas, an Intelligent Reflecting Surface (IRS), and several persons to be localized. The IRS consists of $M$ reflecting elements arranged in a uniform rectangular array (URA) and the separation between the adjacent IRS elements is $d_{irs}$. Each element can reflect the incident signal with an independent phase shift in real-time. The indoor environment area is $d_x \times d_y \times d_z$ meters. For simplicity and without loss of generality, we assume that the persons move horizontally in the X-Y plane. Besides, there are several static reflectors in the region, such as walls and furniture. Let $s_{tx} = e^{-j2\pi f t}$ denote the equivalent complex-valued transmit signal, where $f$ is the carrier frequency in Hertz (Hz).
\subsection{System Model with a Single Reflector}
For the simplicity of analysis, we first consider a simple scenario with only a single reflector. 
As shown in Fig.~\ref{IRS_picture}, there are two propagation paths for the reflector. One is the Tx-reflector-Rx path, in which the signal is transmitted to the object and then reflected to the receiver directly.
The Tx-reflector-Rx propagation attenuation path $\bm{h_{TOR}} \in \mathbb{C}^{ N_r \times 1} $ for the $n$-th receiving antenna can be written as:
\begin{equation}
   \begin{aligned} 
 \relax {h_{TOR}^{n}} & {} = {}{h_{OR}^{n}} \cdot h_{TO} \\
 & {} = {} \rho_{OR}^{n}e^{-j2\pi{ d_{OR}^{n}}/{\lambda} } \cdot \rho_{TO}e^{-j2\pi{d_{TO}}/{\lambda}},
\label{model_H_LOS}
\end{aligned} 
\end{equation}
where $\bm{h_{OR}} \in \mathbb{C}^{N_r \times 1} $ denotes the path response vector from the reflector to the receiver, $ h_{TO} $ is the complex attenuation from the transmitter to the reflector, $\lambda$ is the wavelength of the transmitted signal, $d_{TO}$ is the distance between the transmitter and the object, $d_{OR}^{n}$ is the distance between the reflector and the $n$-th Rx antenna, and ${\rho_{OR}^{n}}$ and ${\rho_{TO}}$ are the corresponding attenuation factor. The other path is the signal transmitted by the Tx reaching the IRS, and then reflected to the moving person, and finally received by the Rx, which can be expressed as 
\begin{equation}
   \begin{aligned} 
  \relax    {h_{TIOR}^{n}}   {} = {}  \rho_{OR}^{n}e^{-j2\pi{d_{OR}^{n}}/{\lambda} } \cdot \sum_{m=1}^{M} \rho_{IO}^{m}\rho_{TI}^{m}  
    \cdot  {e^{-j2\pi({d_{IO}^{m}}+{d_{TI}^{m}})/{\lambda}} } \beta^{m} e^{j\phi^{m}}
\label{model_H_NLOS_total}
\end{aligned} 
\end{equation}
where ${d_{TI}^{m}}$ is the distance from transmitter to the $m$-th element of IRS, $d_{IO}^{m}$ is the distance from the $m$-th element of IRS to the object, $\rho_{IO}^{m}, \rho_{TI}^{m}$ are defined in the similar way as those in (\ref{model_H_LOS}). $\beta^{m} \in [0, 1]$ and $\phi^{m} \in [0, 2\pi)$ denote the amplitude reflection coefficient and phase shift of the $m$-th element, respectively. In this paper, we assume that the IRS only incurs phase shift to the signal, i.e., the amplitude coefficient $\beta$ is 1.  The phase shift of each IRS element is realized by controlling the PIN diodes~\cite{tang2019wireless} or varactor diodes~\cite{li2021high}. The PIN diode can only achieve discrete phase shift, and the number of phase states depends on the number of PIN diodes. The varactor diode can achieve continuous phase control, with the increase of the fabrication cost. 
This article assumes that the IRS adjusts the phase shift by controlling the on or off of the PIN diodes, and denotes the number of quantized bits of the IRS as $B$.  Then, the set of phase states of each IRS element can be expressed as $\mathbb{B} = \{0, \frac{2\pi}{2^{B}}, \dots, \frac{2\pi(2^{B}-1)}{2^{B}} \}$.

Thus, the received signal vector $\bm{s_{rx}} \in \mathbb{C}^{N_r \times 1}$ reflected by a single reflector can be expressed as 
\setlength{\arraycolsep}{0.0em}
\begin{equation}
   \begin{aligned} 
\bm{s_{rx}} & {} = {} ( \bm{h_{TOR}}+ \bm{h_{TIOR}}) \cdot s_{tx} \\
&  {} = {}  (\bm{h_{OR}} h_{TO}+ \bm{h_{OR}}  \bm{h_{IO}}  {\bm{Q}}  \bm{h_{TI}}) \cdot s_{tx}, 
\end{aligned}
\label{single_reflector_signal}
\end{equation}
where $\bm{h_{IO}} \in \mathbb{C}^{1 \times M}$ is the channel vector between the IRS and the reflector, $\bm{h_{TI}} \in \mathbb{C}^{M \times 1}$ is that between the IRS and the transmitter. Let $\bm{q} = [e^{j\phi_{1}}, \dots,  e^{j\phi_{M}}]^{T}$ denote the phase shift vector of the IRS, and  $\bm{Q} = \mathrm{diag}\{\bm{q}\}$ is the weighting matrix of the IRS.

\begin{figure*}[tb]
\centering
	\subfloat[]{
		\includegraphics[width=0.322\textwidth]{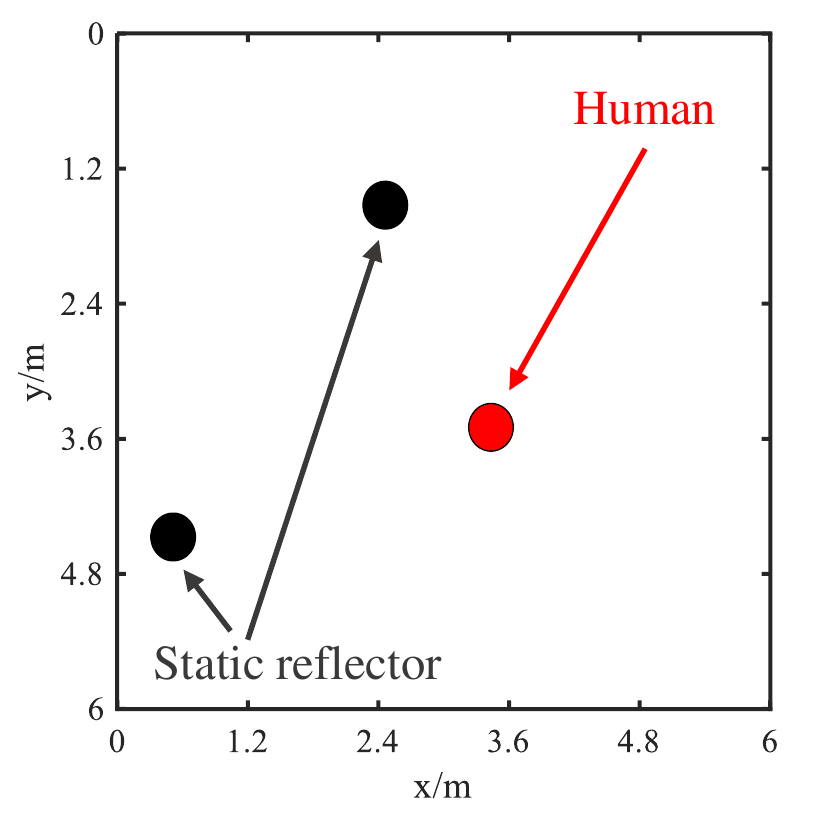}
	}
	\subfloat[]{
		\includegraphics[width=0.315\textwidth]{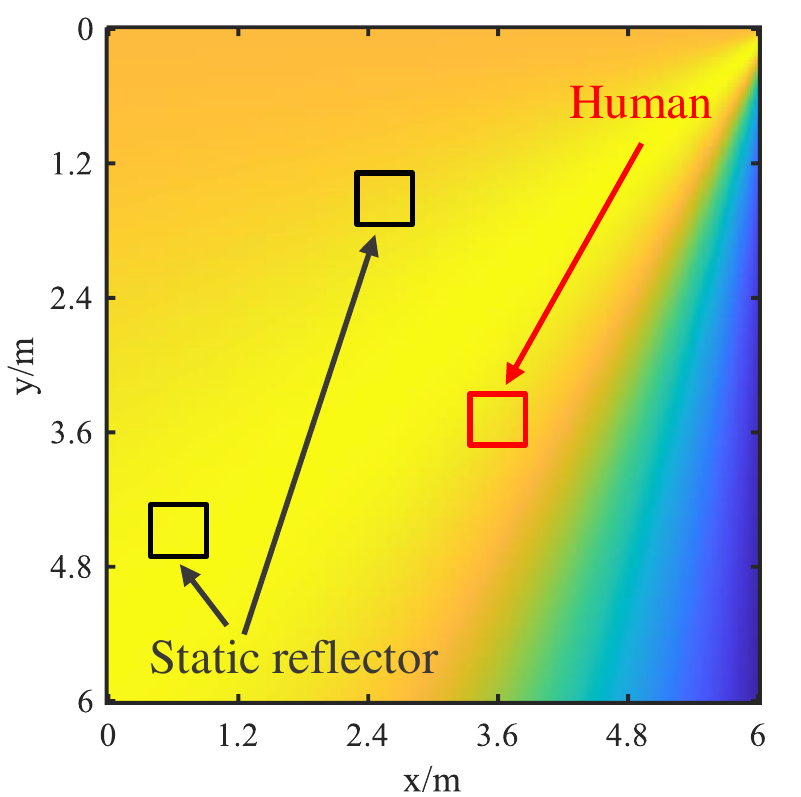}
		}
	\subfloat[]{
		\includegraphics[width=0.315\textwidth]{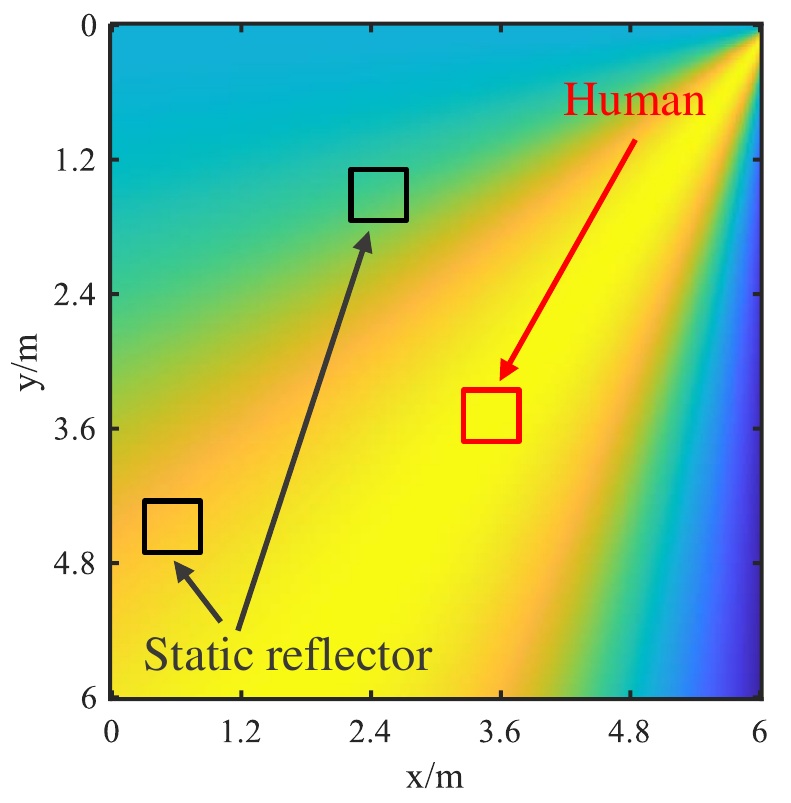}
	}
\caption{
An illustration of the interference of static reflected signals on the passive human localization: (a) the locations of the human and the static reflectors; (b) the signals reflected by static objects (marked with black boxes) and the moving person (marked by the red box), of which the signal reflected from the moving person is completely submerged by the interference; (c) after the subtraction process, the signals reflected from static objects are suppressed and thus the signals reflected from the moving person dominates. 
}
\label{signal_submerged}
\end{figure*}

\subsection{System Model with Multiple Reflectors}
\label{System Signal}
Next, we consider a more practical scenario with multiple persons and static reflectors. 
Let $N_{1}, N_{2}$ denote the number of the persons and the static reflectors, respectively. In this case, in addition to the reflected signal from the human, the signals from static reflectors will also be received by the Rx, which will interfere with our task. The signal propagation paths also include the direct Tx-Rx channel (the signal from Rx to Tx directly) and Tx-IRS-Rx channel (the signal from Rx to IRS, and then reflected to the Rx). Due to the serious path loss, we ignore the signal reflected by IRS two or more times. Hence, the received signal $\bm{s_{rx}}$ is the summation of various static signals and human reflection signals, which is given by
\setlength{\arraycolsep}{0.0em}
\begin{equation}
   \begin{aligned} 
\bm{s_{rx}} & {}= {} (\bm{h_{0}} + \sum_{k=1}^{N_{1}}(\bm{h_{TOR}^{k}} + \bm{h_{TIOR}^{k}}))  s_{tx}  + \bm{\epsilon} \\
& {}= {} (\bm{h_{0}} + \sum_{k=1}^{N_{1}}(\bm{h_{OR}^{k}}h_{TO}^{k}+ \bm{h_{OR}^{k}}  \bm{h_{IO}^{k}}  {\bm{Q}}\bm{h_{TI}}) s_{tx}  + \bm{\epsilon},
\label{model_Y}
\end{aligned}
\end{equation}
where $\bm{h_{0}} \cdot s_{tx}$ denotes the signals which are independent of the moving persons, including the direct Tx-Rx signal, the Tx-IRS-Rx signals, and the signals reflected by static objects, while $\bm{\epsilon} \sim \mathcal{CN}( 0, \sigma^{2}\bm{I}_{N_r})$ denotes the i.i.d. circularly symmetric complex Gaussian (CSCG) noise vector with zero mean and variance $\sigma^{2}$ at the receiver. The transmitter, receiver, and the IRS are placed at known positions, and our objective is to locate the persons with the aid of the IRS.

\section{Single-Person Localization}
\label{section_Single_Person}
In this section, we consider the single-person scenario, to draw important insight into the optimal phase shift for the IRS. 
In order to estimate the person's position, we first need to extract the incoming signal reflected by the person from the mixed sum of all received signals $\bm{s_{rx}}$.  
However, as mentioned in section \ref{System Signal}, besides the human reflection, the signals caused by the multipath are also received by the Rx, which makes it hard to extract human reflection. 
What is even worse is that the direct Tx-Rx signal and the reflections of other static reflectors are generally stronger than that of the moving person. 
To address this challenge, we observe that these interferences, which corresponds to $h_{0}$ in (\ref{model_Y}), keep invariant in the time domain. 
Thus, these signals can be removed by simply subtracting the signals in consecutive measurements under the same IRS phase shift $\bm{q}$, which can be expressed as 
\begin{equation}
\bm{s^{\prime}} = \bm{s_{rx,t}} -  \bm{s_{rx,t-1}}.
\label{method_IRS_subtracting}
\end{equation}

While the static interferences have been eliminated, the signals reflected by the moving person are preserved. 
To localize the moving person, past works leverage the beamforming process by combining constructively the signals from different antennas. More specifically, let $\bm{w} = [w_1, \dots, w_{N_r}] \in \mathbb{C}^{1 \times N_r}$ denote the beamforming vector. The signals  from a specific location $(x, y)$ can be combined by compensating the phase shift on different receiving antennas: 
\begin{equation}
s(x, y) = \bm{w_{x,y}s^{\prime}}
\label{method_w_beamforming}
\end{equation}
where $w_{x,y} =  [w^1_{x,y}, \dots, w^{N_r}_{x,y}]$, $w^{n}_{x,y}$ is defined as 
\begin{equation}
w^n_{x,y}= \text{exp}(j2\pi \frac{ (d_{TO}(x,y)+ d_{OR}^{n}(x,y))}{\lambda}).
\label{method_antennas_phase_w_n}
\end{equation}
Then the X-Y plane can be discretized into a spatial grid with $K$ blocks.  The location of the moving person can be denoted by the index of its block. For each block $k$, (\ref{method_w_beamforming}) can be used to calculate the corresponding $s(x_k,y_k)$. 
Then the location of the person can be given by
\begin{equation}
(\hat{x}, \hat{y}) =   \mathop{\arg\max}_{x_k,y_k} \ \ |s(x_k,y_k)|,
\label{method_phase_w_n_location}
\end{equation}
where $|\cdot|$ denotes the absolute value. However, this method does not work well when the number of antennas is small. To illustrate this, we construct a simulation where the Tx has one transmitting antenna and the Rx has three receiving antennas. Fig.~\ref{signal_submerged} (a) shows the locations of the human and the static reflectors, and Fig.~\ref{signal_submerged} (b) is the corresponding signal after two-dimensional beamforming. As previously mentioned, it can be seen that the human signal is completely submerged by static interference. After the subtraction, the signals reflected by static objects are suppressed and thus the signal reflected by the moving person dominates as shown in Fig.~\ref{signal_submerged} (c). However, the accurate localization of the moving person still cannot be obtained. 
The signals in a large area around the person are all superimposed constructively, resulting in similar signal amplitudes. 
This problem is caused by the fact that WiFi-based systems do not have enough spatial resolution. Specifically, the accuracy of positioning mainly relies on estimating two parameters, the angle-of-arrival (AoA) and time-of-flight (ToF) from the wireless signal. The AoA and ToF resolution can be expressed as \cite{mahafza2013radar}
\begin{equation}
\text{R}_{\text{AoA}} = 2 \arcsin (\frac{\lambda}{Nd}) 
\label{AoA}
\end{equation}
and
\begin{equation}
\text{R}_{\text{ToF}} = \frac{1}{2B}
\label{ToF}
\end{equation}
where $N$ is the number of antennas, $d$ is the inter-element space of the antenna array, and $B$ is the signal bandwidth. For commodity WiFi devices, the transmitted signal bandwidth is generally limited to 40MHz. The inter-element distance is half the wavelength, and the number of receiving antennas is less than or equal to three. Thus, the AoA and ToF resolution, according to \eqref{AoA} and \eqref{ToF}, are 83.6\textdegree \ and 12.5 ns, which means the system does not have enough AoA and ToF resolution. Thus, this method faces fundamental difficulties in accurately estimating the AoA and ToF to obtain accurate localization. On the other hand, IRS has a large number of low-cost controllable elements, which could achieve much higher spatial resolution. With the aid of IRS, the beamforming process at the multi-antenna receiver and the passive phase shift at the IRS can be jointly designed for passive human localization. Intuitively, by jointly optimizing the receiver and IRS, a better localization accuracy could be achieved. 

Our goal is to control the phase shift $\bm{q}$ of the IRS and the beamforming vector $\bm{w}$ to strengthen the signals reflected by the person while keeping the signals surrounding not enhanced.
However, since we cannot know the location of the person in advance, and there is a huge dimension of configurations of the IRS, it is difficult to directly design the best phase shift matrix.
Hence, we adjust the phase shift of IRS to scan the X-Y plane point-by-point to achieve precise positioning.
Compared with mechanical scanning, electrical scanning can achieve a faster data acquisition rate. When the target is located in the main radiation area, the received signal power will be maximized.
On the contrary, if the target is not at the main radiation region, the signals reflected from the target cannot be constructively combined.
In this way, we can increase the amplitude difference of the reflected signal between the target position and its surrounding areas, thereby improving the spatial resolution. Without losing generality, we assume the person is located at $(x,y)$. After the subtraction, the reserved signal can be approximately expressed as $(\bm{h_{TOR}}+ \bm{h_{TIOR}}) \cdot s_{tx} + \bm{\epsilon}$.

\begin{figure}[tb]
		\label{alg:single_person}
		\begin{algorithm}[H]
			\caption{Localization Algorithm for Single-Person Scenario}
			\begin{algorithmic}[1]
				\STATE  Discretize the entire X-Y space into $\mathcal{K}= \{1,...,K \}$.
				\FOR{$k=1 \to K$} 
				\STATE Solve problem (P1) according to \eqref{method_antennas_phase_w_n} and denote the optimal beamforming vector $\bm{w}^{k}$.
				\STATE Solve problem (P2) according to \eqref{method_optimal_phase} and denote the optimal beamforming vector $\bm{q}^{k}$.
				\STATE Measure the signals based on $\bm{w}^{k}$ and $\bm{q}^{k}$, subtract the signals in consecutive measurements.
				\ENDFOR
				\STATE Obtain the location according to \eqref{method_block_k*}.
			\end{algorithmic}
		\end{algorithm}
	\end{figure}
	
The considered optimization problem can be formulated as
\begin{equation}
  \begin{aligned}
  &\mathop{\max}_{\bm{q_{x,y}}, \bm{w_{x,y}}} \ \ \ |\bm{w_{x,y}}(\bm{h_{OR}}h_{TO}+ \bm{h_{OR}} \bm{h_{IO}} {\bm{Q_{x,y}}}  \bm{h_{TI}})|, \\
  & \ \ \ \text{ s.t. } \ \ \ \ \ \  |{q}^{m}_{x,y}| =  \ 1, \ \ \  \forall \  m = 1, \  \dots, \ M, \\
  & \ \ \  \  \ \ \ \ \ \  \ \ \ \  |{w}^{n}_{x,y}| =  \ 1, \ \ \ \forall \  n = 1, \ \dots, \  N_r.
  \end{aligned}
\label{method_single_goal}
\end{equation}
where $\bm{w_{x,y}}$ is the beamforming vector of receiver for the location $(x,y)$, $\bm{q_{x,y}}$ is the phase shift vector of IRS for the location $(x,y)$, and $\bm{Q_{x,y}} = \mathrm{diag}\{ \bm{q_{x,y}}\}$.
In this optimization problem, both the $\bm{w_{x,y}}\bm{h_{OR}}$ and $h_{TO}+ \bm{h_{IO}} {\bm{Q_{x,y}}} \bm{h_{TI}}$ are scalar. Since scalar operation satisfies $|AB| = |A||B|$, the above problem can be decomposed into two sub-problems:
\begin{equation}
  \begin{aligned}
  \mathop{(P1)} \ \ &\mathop{\max}_{\bm{w_{x,y}}} \ \ \ |\bm{w_{x,y}}\bm{h_{OR}}|, \\
  &  \ \mathrm{ s.t. } \ \ \ \ |{w}^{n}_{x,y}| =  \ 1, \ \ \ \forall \   n = 1, \  \dots, \ N_r,
  \end{aligned}
\label{method_questionP1}
\end{equation}
and
\begin{equation}
  \begin{aligned}
  \mathop{(P2)}  \ \ &\mathop{\max}_{\bm{q_{x,y}}} \ \ \ |h_{TO}+ \bm{h_{IO}} {\bm{Q_{x,y}}}  \bm{h_{TI}}|, \\
  & \ \mathrm{ s.t. } \ \ \ \  |{q}^{m}_{x,y}| =  \ 1, \ \ \ \forall  \ m = 1, \  \dots, \ M.
  \end{aligned}
\label{method_questionP2}
\end{equation}

As discussed above, the optimal solution $\bm{{\hat{w}_{x,y}}}$ for sub-problem (P1) is (\ref{method_antennas_phase_w_n}). Sub-problem (P2) can be rewritten as
\begin{equation}
  \begin{aligned} 
  &\mathop{\max}_{\bm{q_{x,y}}} \ \ \ | \rho_{TO}e^{-j2\pi\frac{d_{TO}}{\lambda}} + \sum_{m=1}^{M} \rho_{IO}^{m}\rho_{TI}^{m}{e^{j(\phi^{m}_{x,y} - \frac{2\pi({d_{IO}^{m}}+{d_{TI}^{m}})}{\lambda} }) }|  \\
  &  \ \mathrm{ s.t. } \ \ \  {\phi}^{m}_{x,y} \in [0, 2\pi], \ \ \  \forall \ m = 1, \dots, M.
  \end{aligned}
\label{method_questionP2_rewrite}
\end{equation}
where $q_{x,y}^{m} = e^{j\phi^{m}_{x,y}}$.
According to the Triangle Inequality, we can get
\begin{equation}
  \begin{aligned}
    & \ \ \| \rho_{TO}e^{-j2\pi\frac{d_{TO}}{\lambda}} + \sum_{m=1}^{M} \rho_{IO}^{m}\rho_{TI}^{m}{e^{j({\phi}^{m}_{x,y} - \frac{2\pi({d_{IO}^{m}}+{d_{TI}^{m}})}{\lambda} }) }| \\
    & \leq  |\rho_{TO}e^{-j2\pi\frac{d_{TO}}{\lambda}}| + \sum_{m=1}^{M} |\rho_{IO}^{m}\rho_{TI}^{m}{e^{j({\phi}^{m}_{x,y} - \frac{2\pi({d_{IO}^{m}}+{d_{TI}^{m}})}{\lambda} }) }|.
  \end{aligned}
\label{method_inequality}
\end{equation}

The equality of \eqref{method_inequality} holds if and only if the signals of all different paths are phase-aligned. The corresponding optimal solution $\bm{q^{*}}$ is
\begin{equation}
{\phi}^{*,m}_{x,y}= \frac{2\pi (d_{IO}^{m}+{d_{TI}^{m}-d_{TO}})}{\lambda}, \ \  m = 1, \dots, M.
\label{method_optimal_phase}
\end{equation}

Therefore, the optimal solution for sub-problem (P2) is equivalent to compensate for the phase shift difference caused by the distance difference between $\bm{h_{TOR}}$ and $\bm{h_{TIOR}}$. In this way, the signals from the particular location $(x,y)$ are coherently superimposed. On the contrary, if the reflector is not at the $(x,y)$ position and $\bm{{q^{*}}}$, $\bm{{w^{*}}}$ is still calculated based on the $(x,y)$ position, in this case, the signals reflected from the reflector can not be constructively combined.
Different from the receiver with only a small number of antennas, the IRS consists of a large number of passive elements, which formulates an array with large aperture, leading to a better spatial resolution to distinguish signals from different positions. 
The detailed localization algorithm is shown in Algorithm 1 and described as follows. First, we discretize the entire X-Y space into $\mathcal{K} = \{1, \dots, K \}$ blocks. For each block $(x_k, y_k)$, we calculate the corresponding phase coding pattern $\bm{{q}^{*}_{ x_k, y_k}}$ and beamforming vector $\bm{{w}^{*}_{x_k, y_k}}$ to the main angular radiation region $(x_k, y_k)$. Due to the limited quantization bits of the IRS, we quantize the optimal phase $\bm{{q}^{*}_{x_k, y_k}}$ to their nearest discrete values. Then, we quickly scan across the corresponding beams and collect the signals $s(x_k, y_k)$ for each block. After subtraction to remove the static interference, the human location can be estimated by
\begin{equation}
  \begin{aligned}
(\hat{x}, \hat{y}) &= \mathop{\arg\max}_{x_k, y_k} \ \  (|s(x_k, y_k)|) .
\end{aligned}
\label{method_block_k*}
\end{equation}

\section{Multi-Person Localization}
\label{section_Multi_Person}
In this section, we consider the multi-person scenario.  
Besides the static interference mentioned in section~\ref{section_Single_Person}, another challenge we need to address is the near-far effect in the multi-person scenario. 
Due to the distance difference, the signal energy reflected by the distant persons can be much weaker than that of the near persons, which makes the reflections of distant targets be blurred and difficult to be detected. 
To address this problem, \cite{adib2015multi,zhang2020mtrack} first locate the nearest person and then remove the power in the detected person's location, so as to eliminate the influence of the detected person before proceeding to localize the weaker reflections. 
However, due to the poor spatial resolution, the reflected signals from the sidelobe can be even stronger than that from the mainlobe~\cite{wahlgren2018sidelobe}, leading to the weak reflected target undetected. Thus, the near-far effect cannot be resolved by the aforementioned methods. 

\begin{figure}[!t]
		\label{alg:multi_person}
		\begin{algorithm}[H]
			\caption{Side-lobe Cancellation Algorithm}
			\begin{algorithmic}[1]
				\STATE Initialize the set of detected persons $\mathcal{A}= \{\varnothing\}$, discretize the entire X-Y space into $\mathcal{K}= \{1,...,K \}$ blocks, and set the iteration number $r=1$.
				\STATE Obtain the nearest person's location $(x_{1}, y_{1})$ based on Algorithm 1, and add   $(x_{1}, y_{1})$ to the set $\mathcal{A}$.
				\WHILE {True} 
				    \STATE Update $r=r+1$
					\FOR{$k=1 \to K$} 
				        \STATE Initialize the $\bm{w^{k}}$ and $\bm{q_0^{k}}$ according to  \eqref{method_antennas_phase_w_n} and  \eqref{method_optimal_phase}, and initialize the ${\Delta \bm{\phi}} = [0,   \dots,  0]$.
				        \STATE With the given $\bm{w^{k}}$ and $\bm{q_0^{k}}$, solve problem (P3) and find approximate solution ${\Delta \bm{\phi}}$.
				        \STATE Let ${\Delta \bm{q}} = [e^{j\Delta{\phi}_{1}}, \dots,  e^{j\Delta{\phi}_{M}}]$.
				        \STATE Let $\bm{q^{k}} = \bm{q_0^{k}} \cdot {\Delta \bm{q}}$.
				        \STATE Measure the signal $s(x_{k},y_{k})$ based on $\bm{w}^{k}$ and $\bm{q^{k}}$, subtract the signals in consecutive measurements.
				\ENDFOR
				\STATE Set a square mask centered on the position of the detected persons $\mathcal{A}$, zero all signals in the mask.
				\STATE Select the largest signal $s(x_{r},y_{r})$ among all blocks.
				\IF {The signal energy $s(x_{r},y_{r})$  is lower than the noise floor}
				\STATE  {Break}
				\ENDIF
				\STATE Add $(x_{r}, y_{r})$ to the set $\mathcal{A}$.
				\ENDWHILE
			\end{algorithmic}
		\end{algorithm}
	\end{figure}



To resolve this problem, we have noted that the controllable phase shift of IRS makes it possible to modify the spatial distribution of signals reflection. 
By smartly adjusting the phase shift, the signals $\bm{h_{TOR}}$ and $\bm{h_{TIOR}}$ from the nearby persons can be destructively combined to eliminate their influence on the received signal. Furthermore, the optimization of the phase shift $\bm{q}$ also needs to meet the goal as mentioned in section \ref{section_Single_Person}: the design of the phase shift control should spotlight on a specific block ($x_k, y_k$) to judge whether there is a person present at that location.  

In order to meet these two requirements, we propose a Side-lobe Cancellation algorithm as demonstrated in Algorithm 2. The algorithm is performed in an iterative manner. Specifically, we locate the position of the person with the strongest reflection in each iteration and then eliminate the signal of the detected persons. To achieve this, we go through a two-step process: the first step is optimizing the IRS phase shift $\bm{q^{*}}$ to eliminate the side-lobe interference from the detected persons; the second step is to set a square mask centered on the detected person and zero all signals in the mask. The side length of the square mask is the cross-range resolution for 2-D IRS. Then, the Side-lobe Cancellation algorithm re-scans the spatial grid to find the signal from the next person. We repeat the above procedures iteratively until all the moving persons have been localized.

\begin{figure}[tb]
\centering
	\subfloat[]{
		\includegraphics[width=0.23\textwidth]{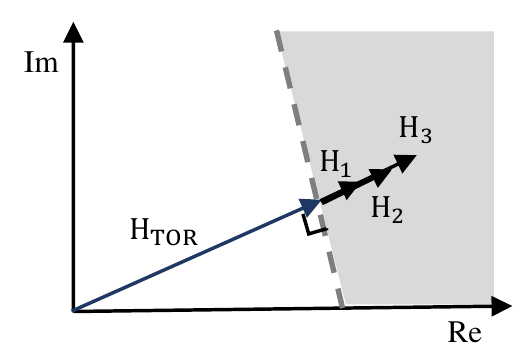}
		}
	\subfloat[]{
		\includegraphics[width=0.23\textwidth]{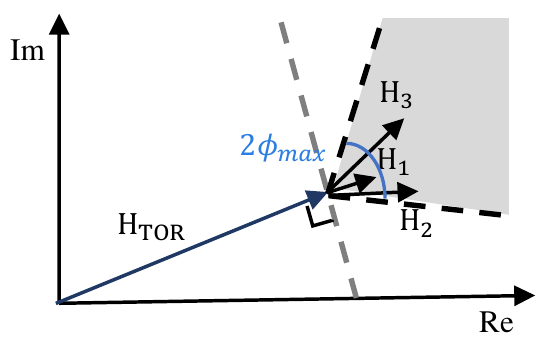}
	}
\caption{ The illustration of the phase shift on the multi-person system: (a) To maximize the signal reflected by the distant person, we first initialize the phase shift $\bm{q_0}$ to align the phases of all signal terms; (b) We optimize the phase perturbation ${\Delta \bm{q}}$ to eliminate the impact of the detected persons. The range of ${\Delta \bm{q}}$ is bounded by $[-\phi_{max}, \phi_{max}]$, as shown in the shaded region, to limit the magnitude of the phase change.}
\label{fig_angle_change}
\end{figure}

Now, we describe the optimization of phase shift in detail. To maintain the goal of \eqref{method_single_goal}, for each block $(x_k, y_k)$,  we calculate the initialization of the phase shift based on the \eqref{method_optimal_phase}, and then find $M$ excitation phase perturbation to minimize the signal power of the detected human. Let $\bm{q_0} \in \mathbb{C}^{1 \times M} = [e^{j\phi_{1}^{0}}, \dots,  e^{j\phi_{M}^{0}}]$, and ${\Delta \bm{q}} \in \mathbb{C}^{1 \times M} = [e^{j\Delta{\phi}_{1}}, \dots,  e^{j\Delta{\phi}_{M}}]$ denote the initial phase vector and phase disturbance vector, respectively. 
Without loss of generality, we assume that it has localized $N$ persons after $N$ iterations.  The signals from the detected persons can be expressed as 
\begin{equation}
\begin{aligned}
    s_d = \mathop{\sum}_{n=1}^{N} \bm{{w^{*}}}\bm{h_{OR}^{n}}(h_{TO}^{n}+ \bm{h_{IO}^{n} (Q_0 } \bm{Q}_{\Delta \bm{q}}) \bm{h_{TI}}),
\end{aligned}
\label{method_multiper_infer}
\end{equation}
where $\bm{{w^{*}}}$ is given by \eqref{method_antennas_phase_w_n}, $\bm{Q_0} = \text{diag} \{ \bm{q_0} \}$ denotes the diagonal matrix whose diagonal elements are the corresponding value of the vector $\bm{q_0}$, and $\bm{Q}_{\Delta \bm{q}} = \text{diag} \{ {\Delta \bm{q}} \}$. Then, the problem can be formulated as follows
\begin{equation}
\begin{aligned}
 \text{(P3)} \ \  &\min_{{\Delta \bm{q}}}  \ \ |s_d  |^{2} \\
    & \ \text{s.t.} \ \ \   | \Delta{\phi_m} | \leqslant {\phi_{max}}, \ \ \  \forall \   m = 1, \  \dots, \  M.
\end{aligned}
\label{method_goal_multiperson}
\end{equation}
where $\phi_{max}$ is a constant used to limit the magnitude of phase perturbation to ensure that the signal from the block $(x_k, y_k)$ maintains phase alignment (as shown in Fig.~\ref{fig_angle_change}).
The following equality holds 
\begin{equation}
\begin{aligned}
    \bm{h_{IO}} (\bm{Q_0}  \bm{Q}_{\Delta \bm{q})} \bm{h_{TI}} =  (\bm{q_0} \cdot {\Delta \bm{q}})  \text{diag} \{\bm{h_{IO}} \}\bm{h_{TI}},
\end{aligned}
\label{method_change_q}
\end{equation}
where $\cdot$ denotes the Hadamard product.
Since the perturbation is small, we can perform a Taylor expansion of the phase perturbation and retain the first two terms, which can be expressed as $e^{j\Delta{\phi}_{m}} \approx 1 + j\Delta{\phi}_{m}$. Thus, we can obtain 
\begin{equation}
\begin{aligned}
   \bm{q_0} \cdot {\Delta \bm{q}} \approx \bm{q_0} + j{\Delta \bm{\phi}} \cdot \bm{q_0} =  \bm{q_0} + j{\Delta \bm{\phi}} \times \boldsymbol{Q_0}
\end{aligned}
\label{method_taylor_q}
\end{equation}
where  ${\Delta \bm{\phi}} \in \mathbb{C}^{1 \times M}= [\Delta{\phi}_{1}, \  \dots, \ \Delta{\phi}_{M}]$. By substituting \eqref{method_change_q} and \eqref{method_taylor_q} into \eqref{method_goal_multiperson}, we rewrite the optimization problem as follows
\begin{equation}
\begin{aligned}
    &  \min_{{\Delta \bm{\phi}}}  \ \ |h_{d}+ {\Delta \bm{\phi}}\bm{h_{D}}|^2 \\
    & \  \text{s.t.} \ \ \ \ \   | \Delta{\phi_m} | \leqslant {\phi_{max}}, \ \ \  \forall \   m = 1, \dots, M.
\end{aligned}
\label{method_goal_multiperson_new}
\end{equation}
where 
\begin{equation}
\begin{aligned}
  h_{d} = \mathop{\sum}_{n=1}^{N} \bm{{w^{*}}}\bm{h_{OR}^{n}}(h_{TO}^{n}+ \bm{q_0} \text{diag} \{\bm{h_{IO}^{n}}\}\bm{h_{TI}}),
\end{aligned}
\label{method_analyse_A}
\end{equation}
and
\begin{equation}
\begin{aligned}
  \bm{h_{D}} = \mathop{\sum}_{n=1}^{N} j\bm{{w^{*}}}\bm{h_{OR}^{n}} \bm{Q_0} \text{diag} \{\bm{h_{IO}^{n}}\}\bm{h_{TI}}.
\end{aligned}
\label{method_analyse_B}
\end{equation}

\begin{figure}[tb]
\centering
\vspace{+0.3cm}
    \includegraphics[width=0.48\textwidth]{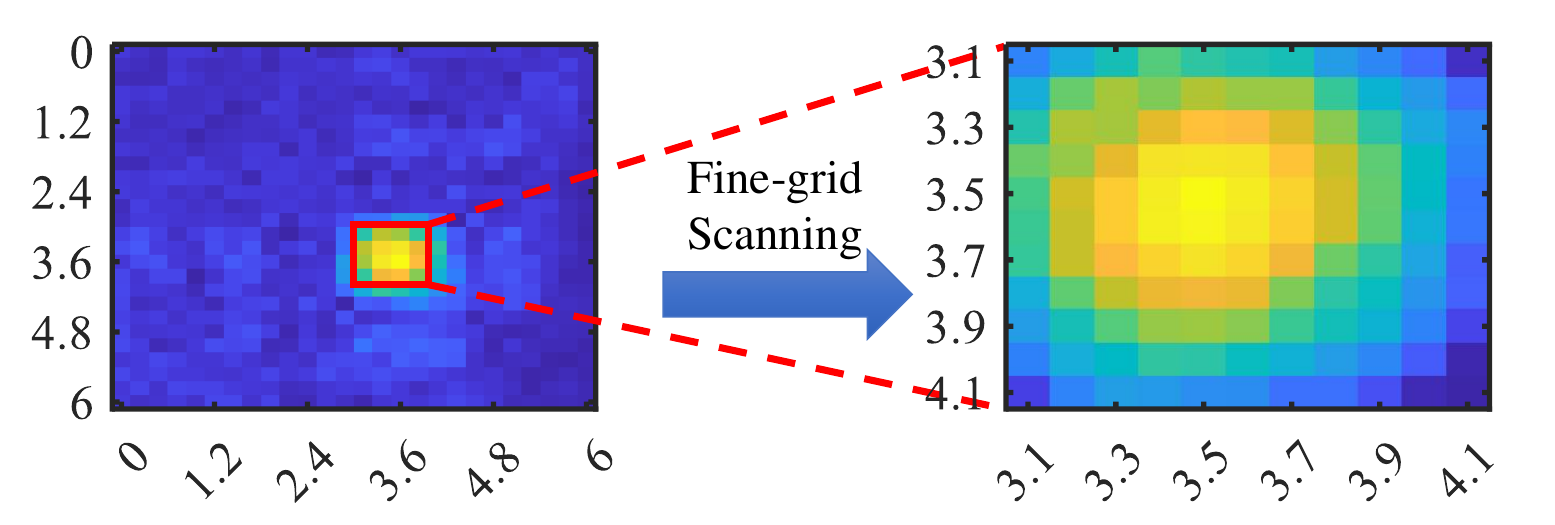}
    \vspace{0.15cm}
\caption{Multi-level localization when the moving target is located at (3.5, 3.5, 0). }
\label{two_step}
\end{figure}

The problem \eqref{method_goal_multiperson_new} is equivalent to 
\begin{equation}
\begin{aligned}
    & \min_{{\Delta \bm{\phi}}}  \ \ ({\Delta \bm{\phi}} \bm{h_{D}}\bm{h_{D}}^{H} {\Delta \bm{\phi}}^{H} + {\Delta \bm{\phi}} \bm{h_{D}}h_{d}^{H} 
     + \ h_{d}\bm{h_{D}}^{H}{\Delta \bm{\phi}}^{H} + h_{d}h_{d}^{H}) \\
    & \  \text{s.t.} \ \ \ \ \   | \Delta{\phi_m} | \leqslant {\phi_{max}}, \ \ \  \forall \   m = 1, \dots, M.
\end{aligned}
\label{method_goal_multiperson_QCQP}
\end{equation}
where the superscript $H$ denotes the conjugate operation. This problem \eqref{method_goal_multiperson_QCQP}  is a quadratically constrained quadratic program (QCQP) which is a fundamental class of non-convex optimization problems. By introducing an auxiliary variable $t$,  it can be reformulated as a homogeneous QCQP problem:
\begin{equation}
\begin{aligned}
    &  \min_{{\Delta \bm{\phi}}}  \ \ \text{tr} (\bm{G}\bm{V}) + h_{d}h_{d}^{H} \\
    & \  \text{s.t.} \ \ \ \ \   \bm{V_{m,m}} \leqslant {\phi_{max}}^{2}, \ \ \  \forall \   m = 1, \dots, M, \\
    &  \ \ \ \ \ \ \ \ \ \bm{V_{M+1,M+1}} = \ 1, \\
    &  \ \ \ \ \ \ \ \ \ \bm{V} \geqslant 0, \ \text{rank}(\bm{V}) = 1.
\end{aligned} 
\label{method_goal_multiperson_QCQPfinal}
\end{equation}
where
\begin{equation}
\begin{aligned}
    \bm{G} = \begin{bmatrix}\bm{h_{D}}\bm{h_{D}}^{H} & \ \   \bm{h_{D}}h_{d}^{H}\\h_{d}\bm{h_{D}}^{H} & \ \   0 \end{bmatrix},
\end{aligned}
\label{method_defination_G}
\end{equation}
and  $\bm{v} = [{\Delta \bm{\phi}}, t] \in \mathbb{C}^{1 \times (M+1)}$, $\bm{V} = \bm{v}^{H}\bm{v}$, $\text{tr}(\bm{G})$, $\text{rank}(\bm{G})$ denote the trace and rank of a matrix $\bm{G}$. Note that the problem \eqref{method_goal_multiperson_QCQPfinal} without the rank-one constraint is a typical convex Semidefinite Programming (SDP) problem, which can be solved by the standard convex optimization algorithms such as the interior-point method. Let $\bm{V}^{*}$ denote the best result of the problem \eqref{method_goal_multiperson_QCQPfinal}. Then, additional steps are required to satisfy the rank-one constraint. After the eigen-decomposition of $\bm{V}^{*}$, $\lambda_{1} \geqslant \lambda_{2} \geqslant \dots\geqslant \lambda_{r} > 0$ are the eigenvalues and $\bm{p_1}, \bm{p_2}, \dots, \bm{p_r}$ are the respective eigenvectors. 
To construct a rank-one solution, we choose $\sqrt{\lambda_{1}}\bm{p_1}$ as our candidate solution to the problem (P3).

Note that we use the Side-lobe Cancellation algorithm to obtain different beam patterns offline. Once finished, we can store it in a lookup table, and later, when we encounter the same pattern, we can directly look up the table to find the corresponding phase shift value.

\begin{figure}[tb]
\centerline{\includegraphics[width=0.49\textwidth]{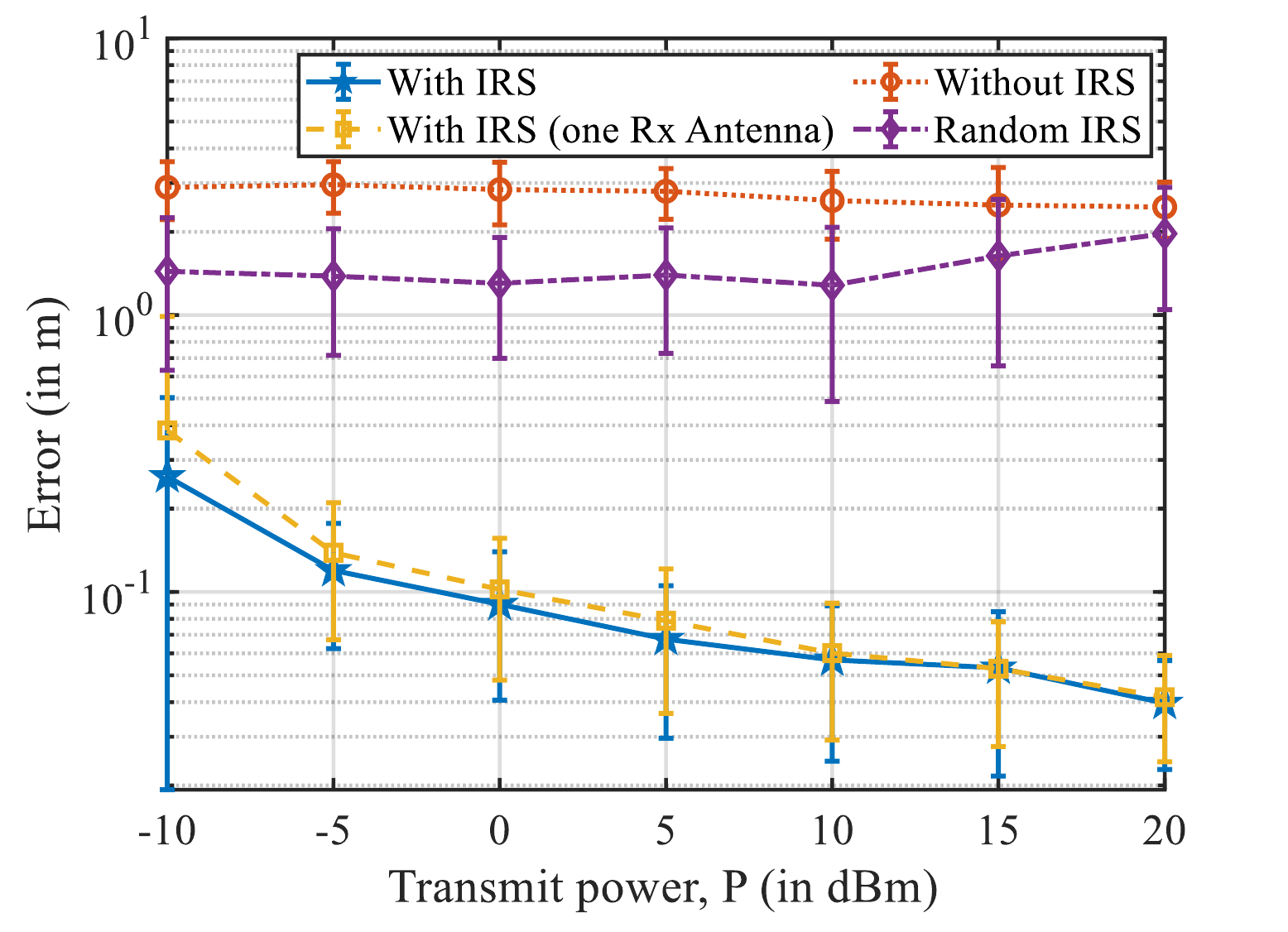}}
\vspace{0.1cm}
\caption{The localization error versus AP power $P$.}
\label{with_withoutIRS}
\end{figure}

\section{Simulation Results}
\label{Simulation}
In this section, we conduct extensive simulations to evaluate the performance of the proposed methods in both the single-person and the multi-person scenarios. 
Under a three-dimensional Cartesian coordinate system,  we assume that the persons move in a room of 6m $\times$ 6m $\times$ 3.5m. The 2-D IRS is located parallel to the X-Y plane with its center at the  (3, 3, 3.5).  The IRS is formulated as a uniform planar array, and the separation space of the IRS element is 0.062m, which is set to approximately half wavelength to avoid the grating lobe. The single-antenna transmitter is located at the  (3, 3, 3), while the three-antenna receiver is placed at (6, 0, 0) and the inter-element space of the antenna array is 0.062m.  We assume that all antennas of Tx and Rx are omnidirectional. The signal frequency is $f$ = 2.4GHz.  We assume there are two static reflectors in the room, placed at (2, 2, 0) and (3, 5.5, 0) respectively. The channel coefficient $\rho$ for all the channels is modeled as 
\begin{equation}
    \rho(d) = \sqrt{\rho_0(\frac{d}{d_0})^{-\alpha}},
\label{result_path_loss}
\end{equation}
where $\rho_0 = -20$dB is the path loss at the reference distance $d_0 = 1 \rm{m}$, $d$ is the individual link distance, $\alpha$ denotes the path loss exponent and represents how fast the power decays with distance, which depends on the propagation environment. By referring to the parameters of ~\cite{zhang2020capacity}, we set $\alpha_{TO} = 3.6$, $\alpha_{TI} = 2.2$, $\alpha_{IO} = 2.2$, $\alpha_{OR} = 3.6$ as the path loss exponents of the Tx-reflector link, the Tx-IRS link, the IRS-reflector link, the reflector-Rx link, respectively. We denote the Gaussian noises at the receiver with mean zero and variance $\sigma^2 = -80$ dBm, which is consistent with the parameter in the ~\cite{zhang2020capacity}.  
We adopt the root mean square error (RMSE) to quantitatively evaluate the accuracy of the localization. The localization error through a Monte Carlo process is defined as:
\begin{equation}
{\rm RMSE} = \frac{1}{T} \begin{matrix} \sum_{t=1}^T \sqrt{(x_{t} - \widehat{x}_t )^{2} + (y_{t} - \widehat{y}_t )^{2}} \end{matrix}
\label{error_function}
\end{equation}
where $(\widehat{x}_t, \widehat{y}_t)$ is the estimated location of the moving person,  $(x_{t}, y_{t})$ is the ground-truth, and $T$ is the number of Monte Carlo runs.

\begin{figure}[tb]
\centerline{\includegraphics[width=0.48\textwidth]{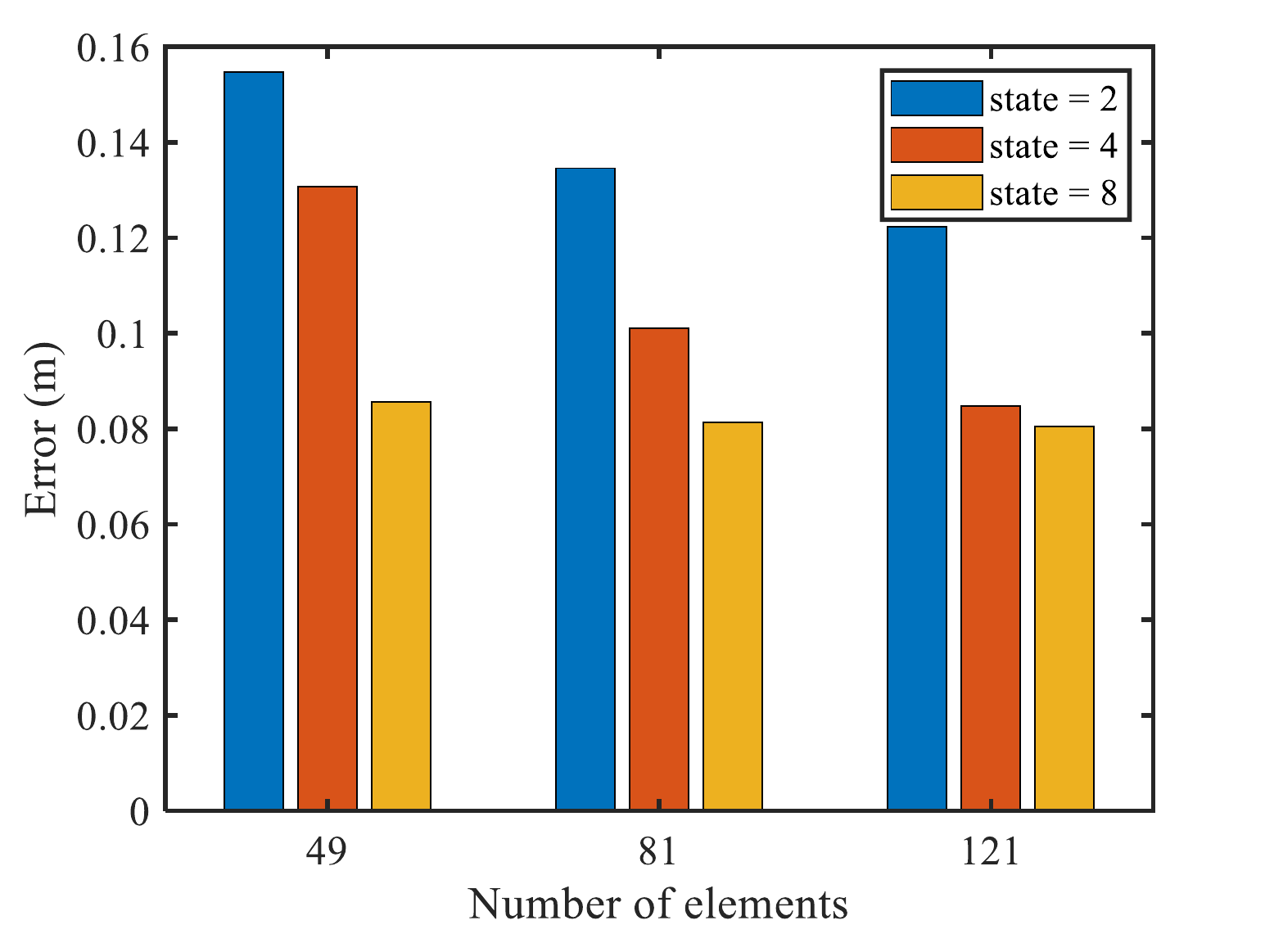}}
\caption{The localization error versus the number of IRS element and the number of states when AP power $P=0$ dBm.}
\label{different_element}
\end{figure}

\subsection{Single-person Scenario}
We start with a single-person scenario to evaluate the positioning accuracy of Algorithm 1.
The person is located at (3.5, 3.5, 0). The IRS consists of 9 $\times$ 9 elements, and each element has four states. Moreover, to reduce the computational complexity,  we adopt a multi-level space division method, which firstly performs coarse segmentation on the X-Y space, and then selects the block with the highest amplitude for fine-grained segmentation, as shown in Fig.~\ref{two_step}. 
We can see that, with the aid of IRS, the proposed method can locate the person accurately. In comparison, the system without IRS in Fig.~\ref{signal_submerged}(c) can only coarsely get the AOA of the person, but cannot accurately estimate the position of the target due to the low spatial resolution of WiFi devices.  These visual comparisons demonstrate that the proposed algorithm can achieve better spatial resolution.

To further demonstrate the superiority of the proposed framework, we compare its performance with the following benchmark schemes:
\begin{enumerate}
\item  \textbf{Without IRS}: It does not utilize the IRS and realizes localization only via the beamforming of the receiving antennas.
\item \textbf{Random IRS}: Randomly feeds the initial phase weight to each IRS unit.
\item \textbf{With IRS (One Rx Antenna)}: the number of receiving antennas decreases to one, which removes the beamforming capability of the receiving antennas.
\end{enumerate}

Fig.~\ref{with_withoutIRS} presents the localization error versus the standard deviation for the proposed method and the benchmark schemes. It is observed that the proposed method significantly improves the accuracy of localization, and outperforms other benchmark schemes. As the comparison algorithms, the RMSE of scheme 1) and scheme 2) are maintained at about 2.8m and 1.6m, which means only using traditional WiFi devices for localization in this scenario does not work.

On the other hand, we can observe that when the AP power is higher than 5 dBm,  scheme 3) has comparable localization accuracy with the three-antenna method. It implies that even if the original WiFi device does not have AoA and ToF resolution, the IRS-aided localization method can still achieve fine-grained positioning accuracy. However, the difference in localization accuracy of these two methods gradually becomes larger as the AP power decreases.  By increasing the number of antennas, the receiver can focus the signal power towards the target location through the beamforming process, thereby alleviating static multipath interference and being more robust to noise.

\begin{figure*}[tb]
	\centering
	\subfloat[]{
		\includegraphics[width=0.313\textwidth]{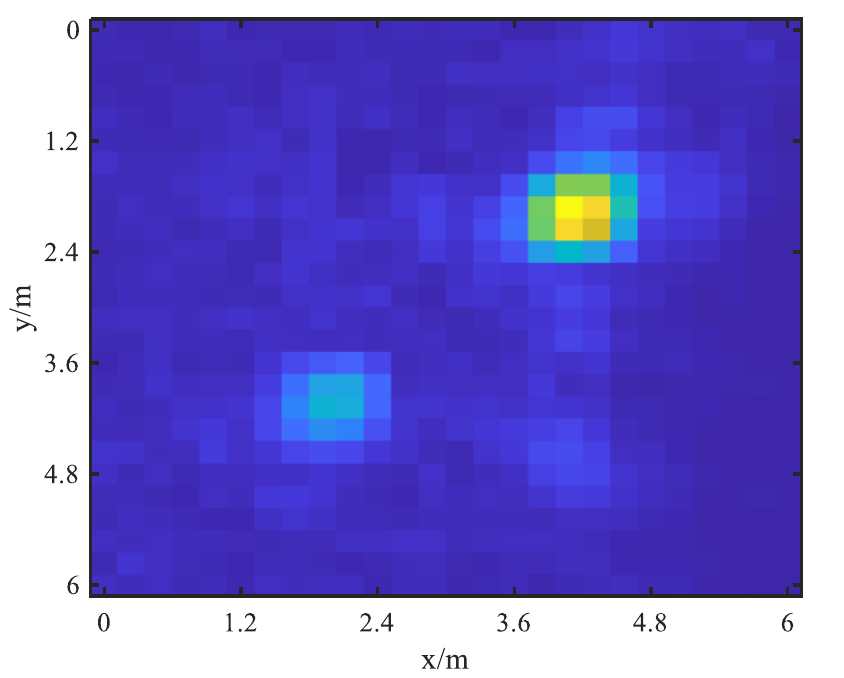}
	}
	\subfloat[]{
		\includegraphics[width=0.313\textwidth]{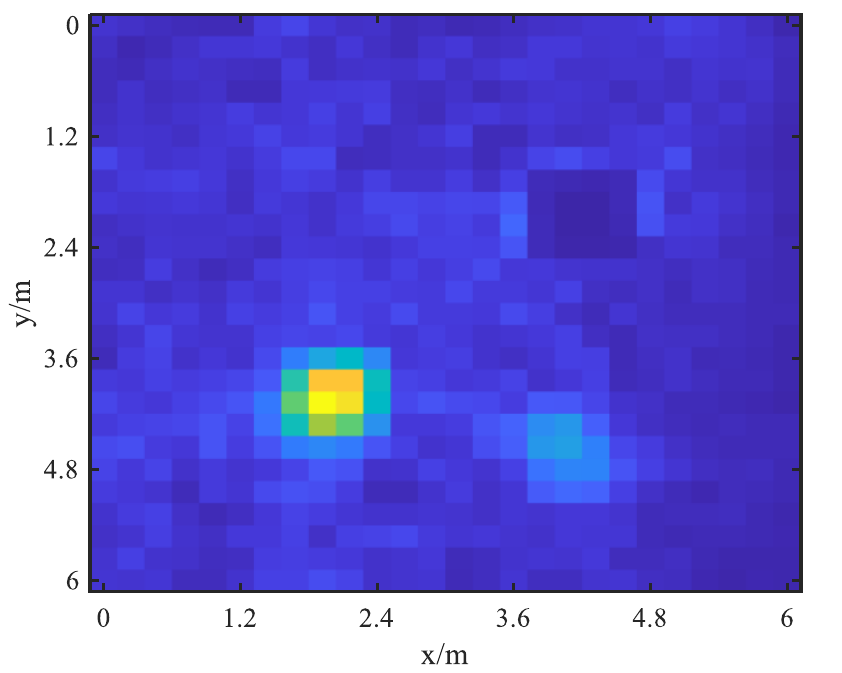}
	}
	\subfloat[]{
		\includegraphics[width=0.313\textwidth]{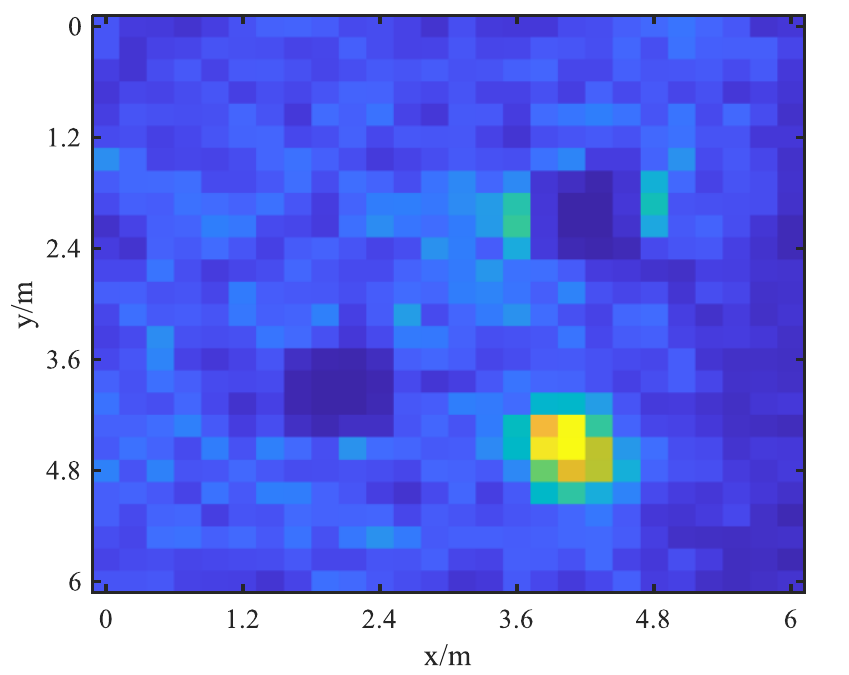}
	}\\
	
	\subfloat[]{
		\includegraphics[width=0.313\textwidth]{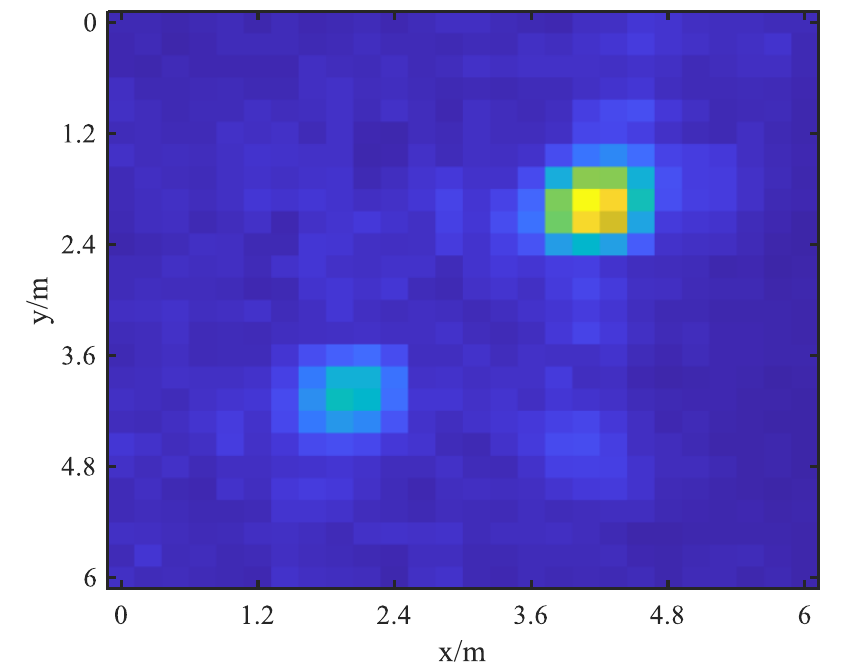}
	}
	\subfloat[]{
		\includegraphics[width=0.313\textwidth]{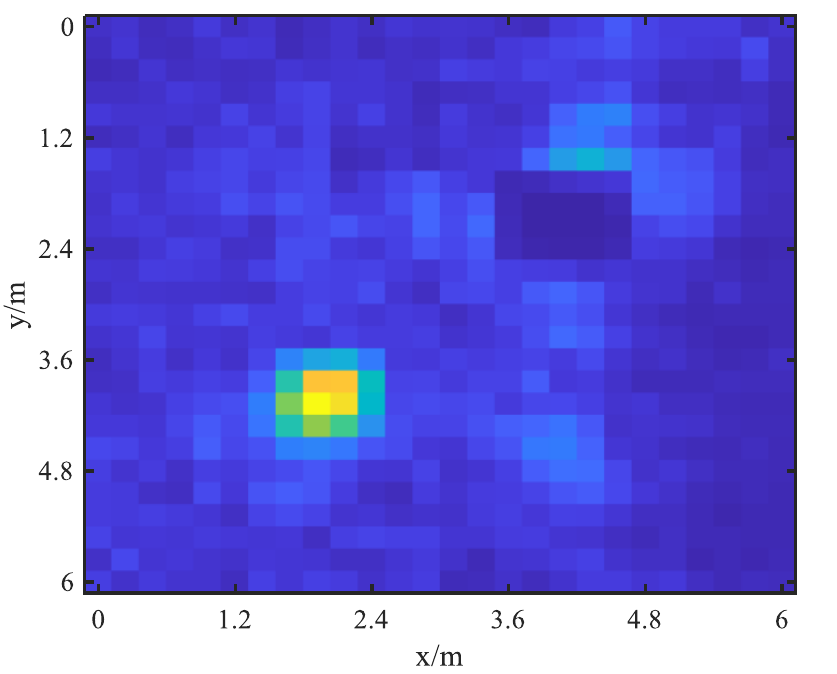}
	}
	\subfloat[]{
		\includegraphics[width=0.313\textwidth]{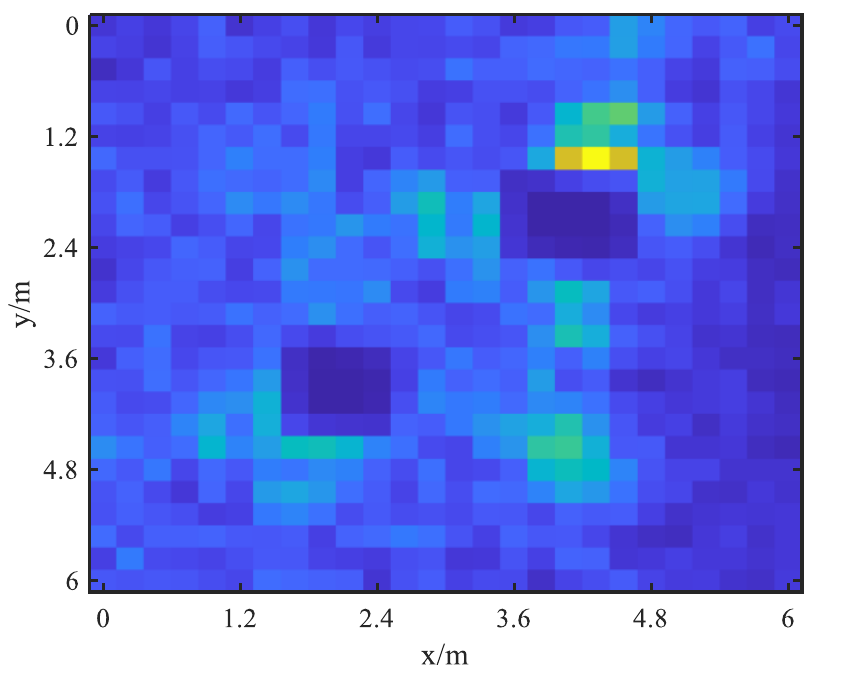}
	}
	\caption{Localization results with different methods in multi-person scenario ($P_{AP}$ = 15 dBm): (a)-(c) show the localization result of the IRS-based Side-lobe Cancellation method. (a) shows the 2D heatmap in the presence of three persons; (b) and (c) show the heatmap obtained after canceling out the first, second person respectively; (d)-(f) show the corresponding results without Side-lobe Cancellation algorithm. }
	\label{fig_multi_compare}
\end{figure*}

We further explore the impact of the number of IRS elements and the number of states. 
Fig.~\ref{different_element} depicts the localization error for different number of IRS elements ($7\times 7,\ 9\times 9,\ 11\times 11 $) and IRS states ($2,\ 4,\ 8$)  when AP power $P=0$ dBm. We can observe that the localization error decreases when the number of IRS elements and IRS states increases, which is in line with expectations. Increasing the number of elements, on the one hand, increases the IRS aperture size and obtains a better spatial resolution, on the other hand, can achieve a constructive combination of more signals and be more robust to noise.  The increase in the number of states can reduce the quantization error. Furthermore, it is interesting to note that, even the IRS has only two-phase states (0 or 180\textdegree), which has been widely implemented~\cite{xu2016dynamical, pei2021ris, li2020intelligent}, the proposed method still has a centimeter-level localization accuracy.

\subsection{Multi-person Scenario}
In the following, we evaluate the proposed algorithm under the multi-person scenario.  We simulate the positioning results in a three-person scenario and the locations for the persons are  (4.1, 2), (2, 4), (4, 4.5), 
respectively. We consider the IRS with 11 $\times$ 11 elements and 8 phase shift states. We set the $\phi_{max}$ = $\frac{\pi}{6}$ to ensure that the approximate error of~\eqref{method_taylor_q} is small. We set the side length of the square mask to be 0.5m, which is approximately equal to the cross-range resolution of the IRS. 
Fig.~\ref{fig_multi_compare} (a)-(c) show the coarse localization heatmaps obtained in the presence of three persons. In Fig.~\ref{fig_multi_compare} (a), we can coarsely estimate the locations of two persons. The person at (4.1, 2) can be clearly seen, and the signal at (2, 4) is relatively weak. The signal of the person at (4, 4.5) is completely overwhelmed by noise and the interference of the other two persons. After fine-grained localization of the person with the strongest reflection 
(4,1, 2), we eliminate the spread of reflected signals from the localized person in all blocks, as shown in Fig.~\ref{fig_multi_compare} (b). Now, the strongest signal is reflected from the position (2, 4), and then we repeat the above procedures. After eliminating the reflected signals of the detected persons, the signal from the person at (4, 4.5) is visible in Fig.~\ref{fig_multi_compare} (c).

\begin{figure*}[tb]
	\centering
	\subfloat[]{
		\includegraphics[width=0.32\textwidth]{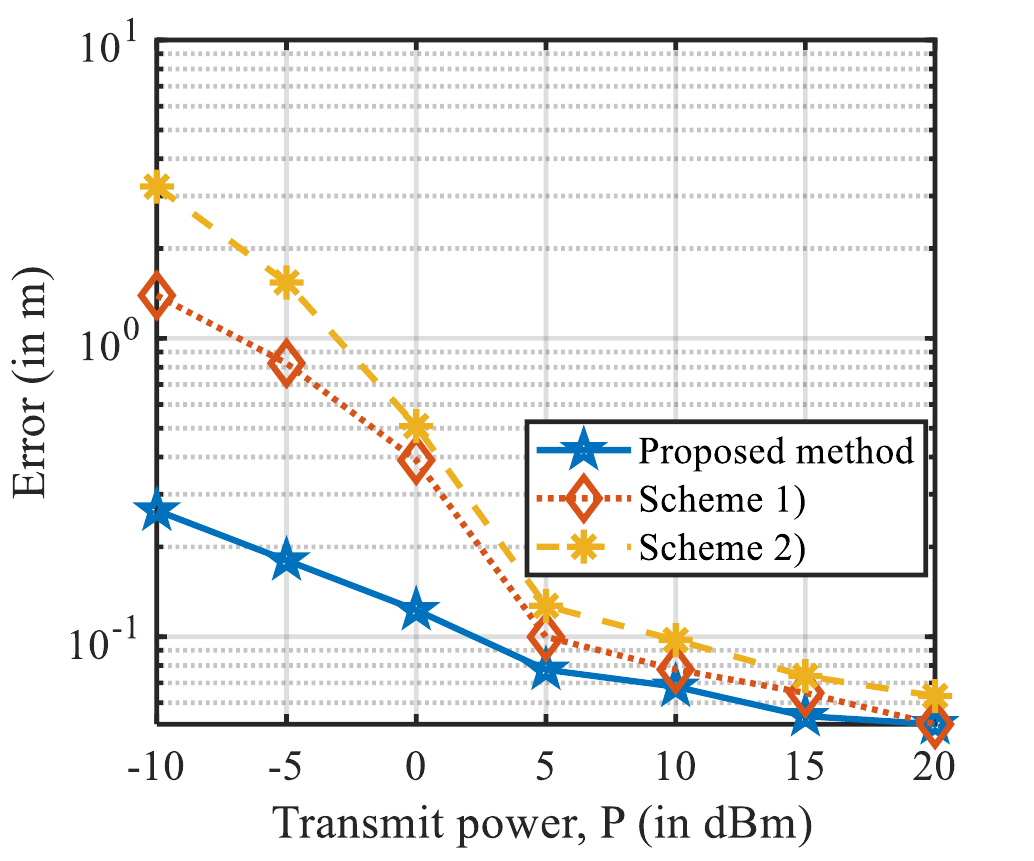}
	}
	\subfloat[]{
		\includegraphics[width=0.32\textwidth]{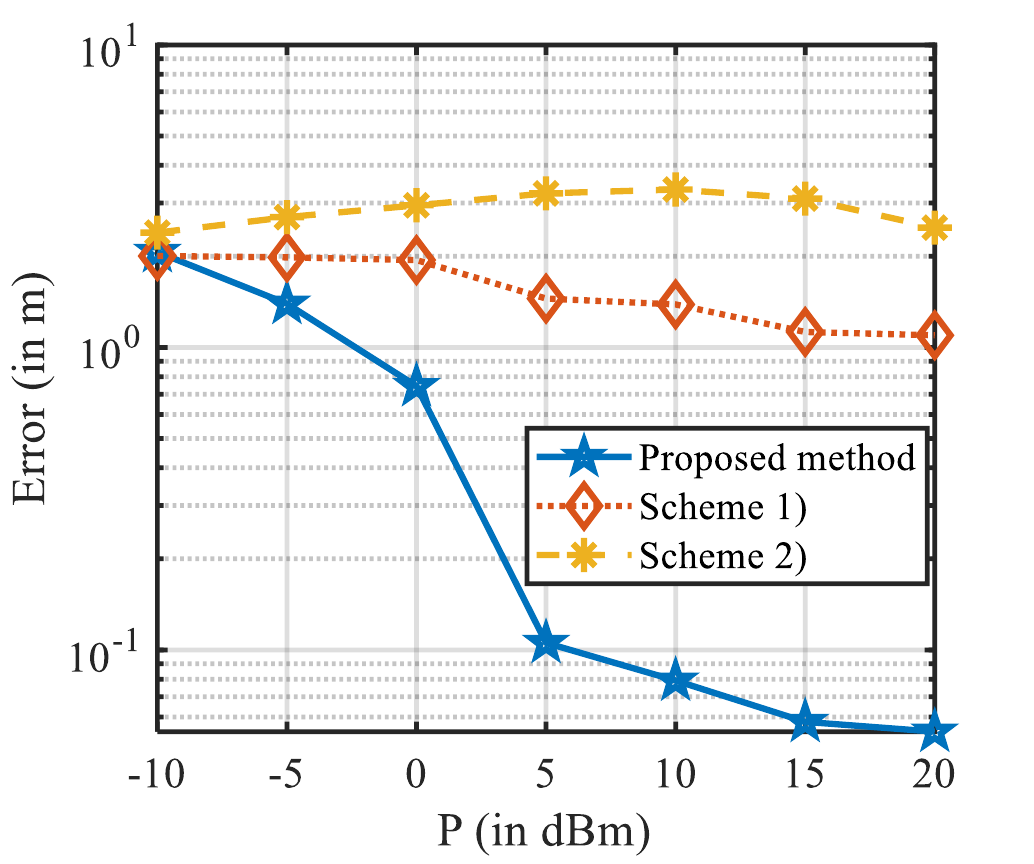}
	}
	\subfloat[]{
		\includegraphics[width=0.32\textwidth]{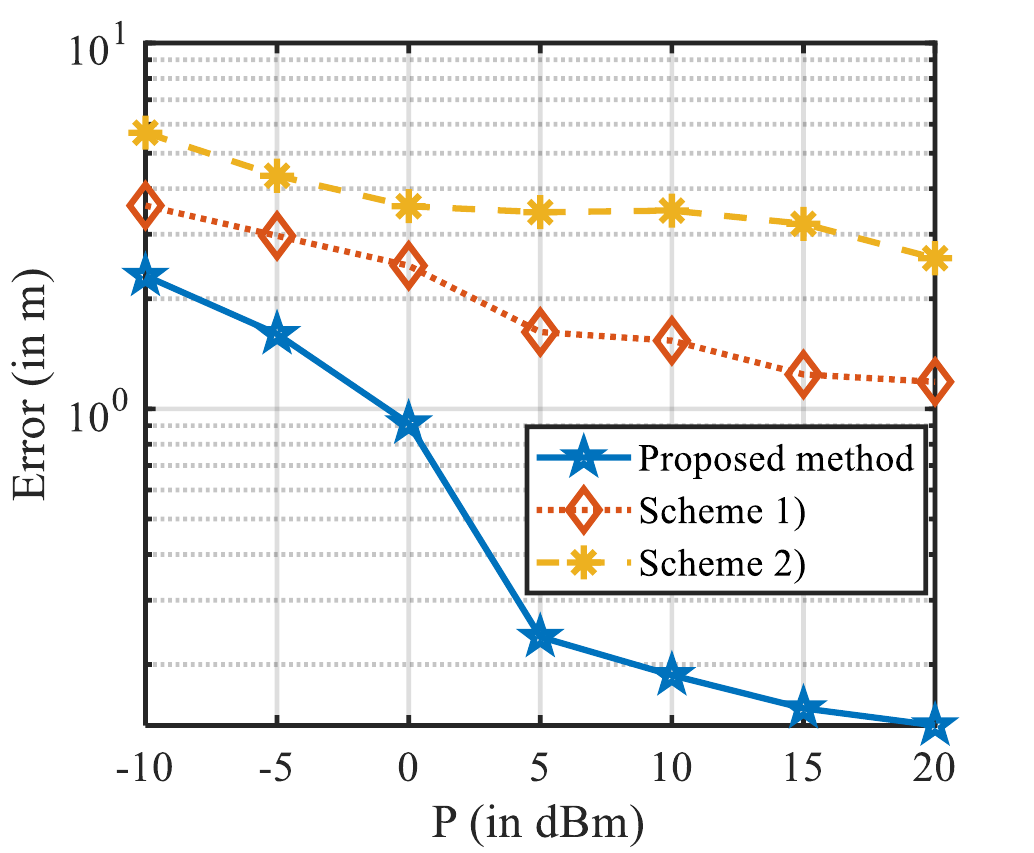}
	}
	\caption{The localization error versus different methods in a three-person scenario: (a) the localization error for the second detected person; (b) the localization error for the third detected person; (c) the sum of the localization errors of the three persons.}
	\label{fig_multi_error_humannumber}
\end{figure*}

\begin{figure*}[tb]
	\centering
	\subfloat[]{
		\includegraphics[width=0.313\textwidth]{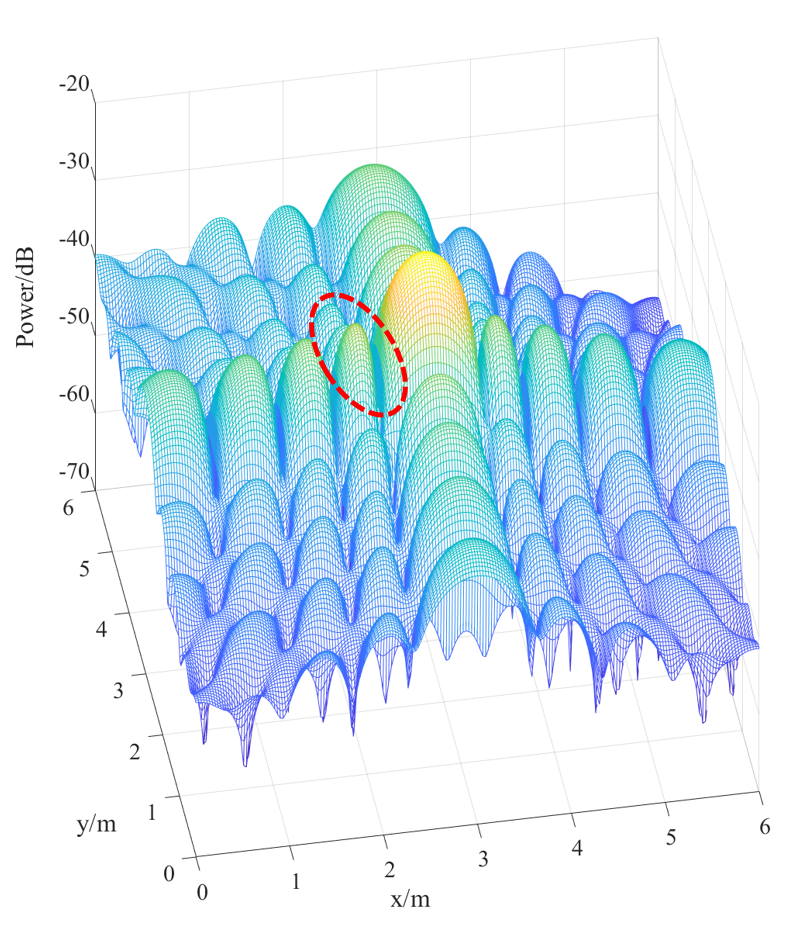}
		}
	\subfloat[]{
		\includegraphics[width=0.313\textwidth]{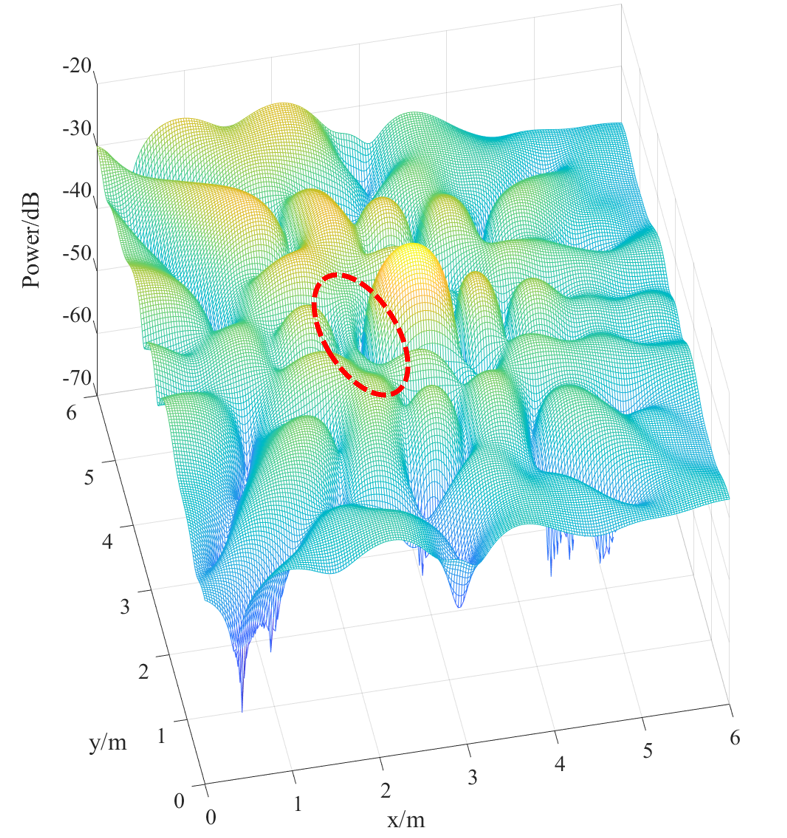}
	}
	\subfloat[]{
		\includegraphics[width=0.313\textwidth]{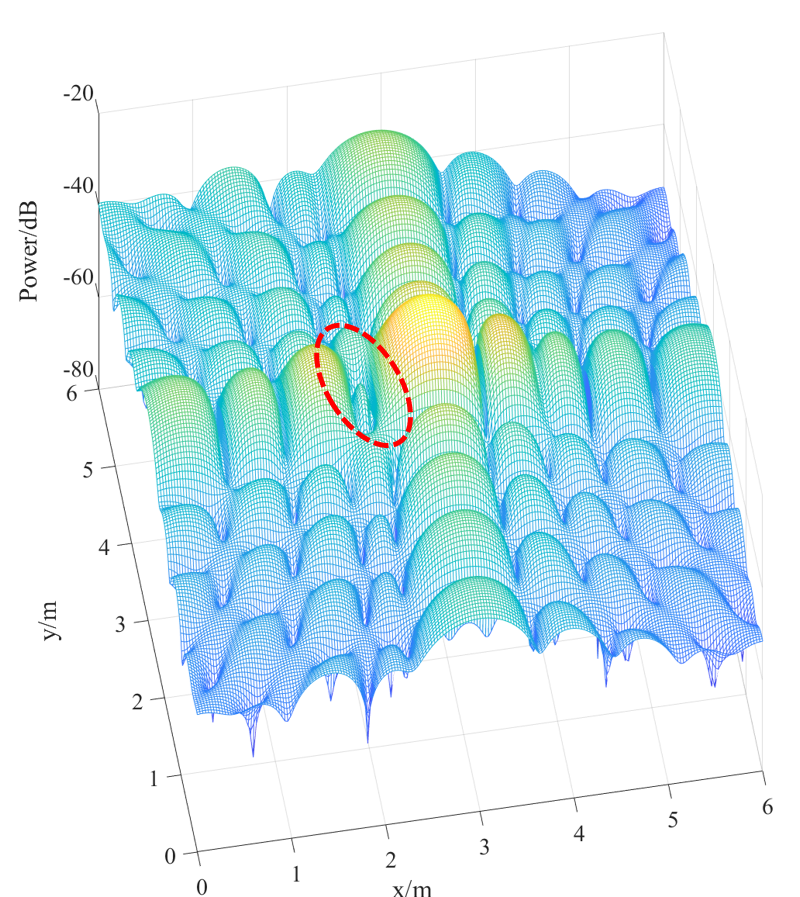}
	}
	\caption{The beam pattern versus different methods in the X-Y plane. We compare the suppression of sidelobes by different algorithms. We set the main lobe points at (3, 3), and the interference at (2.22, 3): (a) The beam pattern of Scheme 1). The sidelobe signal at (2.22, 3) cannot be suppressed; (b) The beam pattern of Scheme 2). The sidelobe signal  is suppressed, but the sidelobe signal at other positions is enhanced, and the main lobe power is reduced; (c) The beam pattern of the proposed scheme. While the sidelobe interference is eliminated, the beam pattern at other locations remains almost unchanged.
	}
	\label{fig_pattern_compare}
\end{figure*}

Next, we compare the performance of the proposed method with the algorithm without Side-lobe Cancellation. Fig.~\ref{fig_multi_compare} (d)-(f) show the results of the comparison method.  In Fig.~\ref{fig_multi_compare} (d), (e), the algorithm without Side-lobe Cancellation can detect the location of target (4,1, 2) and target (2, 4). However, even after zeroing in Fig.~\ref{fig_multi_compare} (f), the power in (4, 4.5) is still weaker than many blocks in space, i.e., it fails to locate the third person. The above results further demonstrate the necessity of jointly optimizing the phase shift of IRS and the zero operation for interference cancellation in the multi-person scenario.

Then, we adopt RMSE to quantitatively evaluate the accuracy of multi-person positioning.  
We compare the performance of the proposed method with the following algorithms:
\begin{enumerate}
\item  \textbf{Without Side-lobe Cancellation}: optimize the IRS with Algorithm 1.
\item   \textbf{Maximum SINR optimization}: replaces the optimization goal to maximizing the SINR ratio. The optimization is formulated as
\begin{equation}
\begin{aligned}
\max_{\bm{q}} \frac{|\bm{w^{*}}\bm{h_{OR}}(h_{TO}+ \bm{h_{IO}Q h_{TI}}|^{2}}{|\mathop{\sum}_{n=1}^{N} \bm{w^{*}}\bm{h_{OR}^{n}}(h_{TO}^{n}+ \bm{h_{IO}^{n} Q h_{TI}}|^{2} + \sigma^{2}}
\end{aligned} 
\label{Maximum_SINR_optimization}
\end{equation}
where the numerator is the signal energy reflected by the target position, and the denominator represents the total energy of the interference signal and noise. We refer to \cite{cui2019secure} to solve this problem.
\end{enumerate}

Since the three schemes have the same locating method for the nearest person, only the location error of the distant persons is compared here.  As shown in Fig.~\ref{fig_multi_error_humannumber}, it is observed that by adding Side-lobe Cancellation, the localization error is significantly reduced for the second person, compared to the case without Side-lobe Cancellation. For the third person with the weakest reflected signal, the other two methods cannot extract the target's signal in the presence of interference and noise, but our method can accurately estimate the location. Furthermore, it is interesting to note that although Scheme 2) reduces interference by maximizing SINR, its result is not even as good as Scheme 1).  In order to analyze this phenomenon, we compared the beam pattern of the three methods. We map the angle pattern to the X-Y plane, as shown in Fig.~\ref{fig_pattern_compare}. We set the main lobe points to (3, 3), and the interference at (2.22, 3). The selected interference point is the position where the side lobes are higher in the initial pattern. This position is chosen to better verify the effectiveness of the three algorithms. Scheme 1) does not perform interference cancellation, thus, the sidelobe energy at position (2.22, 3) is the highest in these three methods. 
Although Algorithm 2) reduces the side-lobe (2.22, 3), it causes the side-lobes of other positions to become larger and the amplitude of the main lobe becomes weaker. Compared with Scheme 1), the IRS-based Side-lobe Cancellation Algorithm significantly reduces 13dB in the sidelobe (2.22, 3), and has little effect on the shape of the beam pattern.


\begin{figure*}[tb]
	\centering
	\subfloat[]{
		\includegraphics[width=0.32\textwidth]{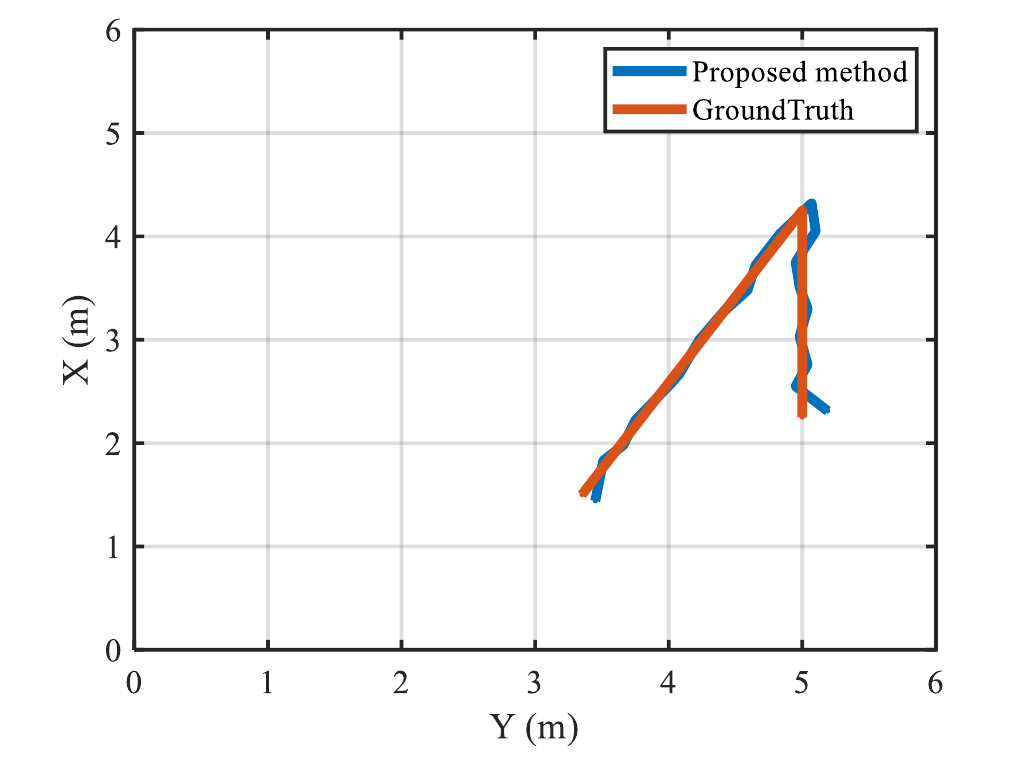}
	}
	\subfloat[]{
		\includegraphics[width=0.32\textwidth]{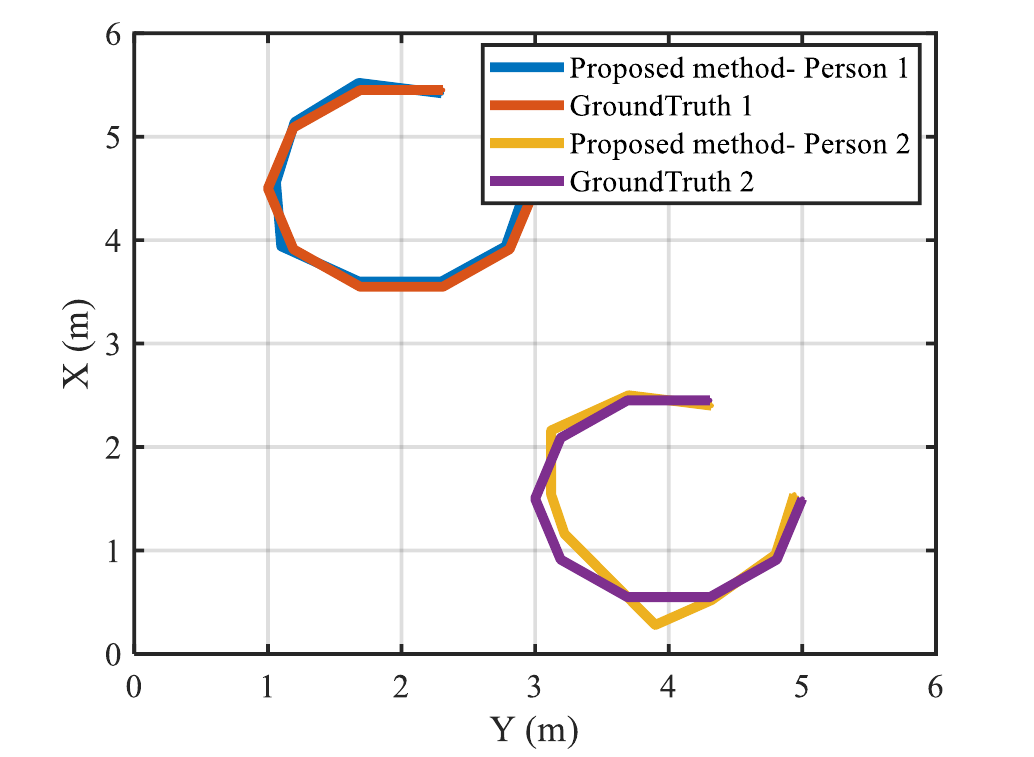}
	}
	\subfloat[]{
		\includegraphics[width=0.32\textwidth]{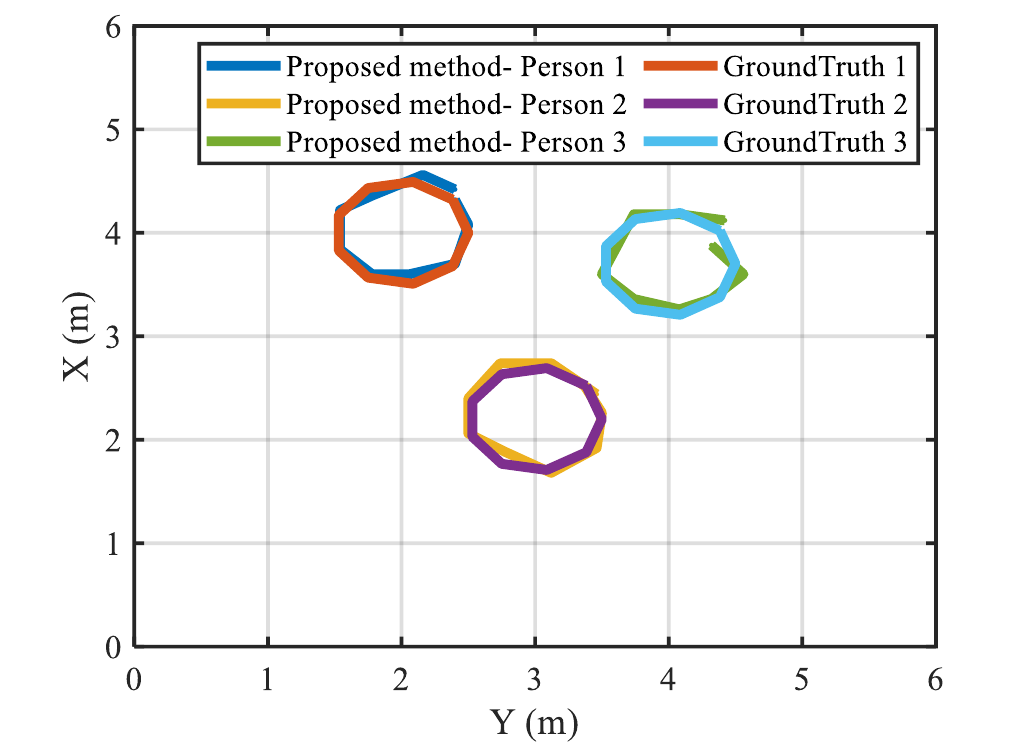}
	}
	\caption{Tracking results of the proposed method in different scenarios when AP power $P= 15$ dBm.}
	\label{IRS_trajectory}
\end{figure*}

In order to determine whether our method consistently works for different scenarios, Fig.~\ref{IRS_trajectory} plots three different estimated trajectories and ground truth in the coordinate system.  It can be seen that our method has achieved accurate positioning and tracking in both the single-person and the multi-person scenarios.

\section{Conclusion}
\label{Conclusion}
In this paper, we investigated to achieve passive human localization with the aid of IRS in both single-person and multi-person scenarios. 
In the single-person scenario, we derived the closed-form solution to optimally control the phase shift of IRS elements. In the multi-person scenario, we proposed a Side-lobe Cancellation algorithm to resolve the near-far effect. Simulation results demonstrated that the proposed method achieved accurate localization of multiple moving persons, with a pair of commodity WiFi devices and IRS. Such an accurate localization performance can only be achieved by expensive dedicated hardware previously. It is worth noting that the method proposed in this paper can also be applicable to other WiFi sensing problems including imaging, gesture recognition, vital sign monitoring, etc.   

\bibliographystyle{IEEEtran}
\bibliography{references}


 





\end{document}